\documentclass[10pt,twoside,english,final, 5p, preprint, times,numbers, sort&compress]{elsarticle}
\usepackage[T1]{fontenc}
\usepackage{amsthm}
\usepackage{amsmath}
\usepackage{amssymb}
\usepackage{graphicx}

\makeatletter

\providecommand{\tabularnewline}{\\}

\numberwithin{equation}{section}
\numberwithin{figure}{section}

\makeatother

\usepackage{babel}
\begin{document}

\title{Extended force density method and its expressions}

\author{Masaaki Miki}

\ead{mikity@iis.u-tokyo.ac.jp}

\address{Department of Architecture, School of Engineering, the University
of Tokyo, komaba4-6-1, meguro-ku, Tokyo, 153-8505, JAPAN}
\begin{abstract}
The objective of this work can be divided into two parts. The first
one is to propose an extension of the force density method (FDM)\citep{Schek74},
a form-finding method for prestressed cable-net structures. The second
one is to present a review of various form-finding methods for tension
structures, in the relation with the extended FDM.

In the first part, it is pointed out that the original FDM become
useless when it is applied to the prestressed structures that consist
of combinations of both tension and compression members, while the
FDM is usually advantageous in form-finding analysis of cable-nets.
To eliminate the limitation, a functional whose stationary problem
simply represents the FDM is firstly proposed. Additionally, the existence
of a variational principle in the FDM is also indicated. Then, the
FDM is extensively redefined by generalizing the formulation of the
functional. As the result, the generalized functionals enable us to
find the forms of tension structures that consist of combinations
of both tension and compression members, such as tensegrities and
suspended membranes with compression struts.

In the second part, it is indicated the important role of three expressions
used by the description of the extended FDM, such as stationary problems
of functionals, the principle of virtual work and stationary conditions
using $\nabla$ symbol. They can be commonly found in general problems
of statics, whereas the original FDM only provides a particular form
of equilibrium equation. Then, to demonstrate the advantage of such
expressions, various form-finding methods are reviewed and compared.
As the result, the common features and the differences over various
form-finding methods are examined. Finally, to give an overview of
the reviewed methods, the corresponding expressions are shown in the
form of three tables.\end{abstract}
\maketitle

\section{Introduction}

This is a revised version of \citep{Miki10}.

The objective of the first half of this work is to propose an extension
of the force density method (FDM)\citep{Schek74}, a form-finding
method for prestressed cable-net structures. Particularly, for the
prestressed tension structures, form-finding is a process to ensure
them to have a prestress state, because the existence of a prestress
state highly depends on the form of the tension structure.

In section 2, the original FDM is described with its major advantage
in form-finding process of cable-net structures. In addition, it is
pointed out that the FDM become useless when it is applied to the
prestressed structures that consist of combinations of both tension
and compression members, e.g. tensegrities. Therefore, the FDM has
a scope for extension. 

In section 3, a functional whose stationary problem simply represents
the original FDM is firstly proposed. Additionally, the existence
of a variational principle in the FDM is also indicated, although
the formulations provided by the original FDM look different from
those related to the variational principle. The clarified functional
enables an extension of the FDM.

In section 4, the FDM is extensively redefined by generalizing the
formulation of the functional. As the result, the generalized functionals
enable us to find the forms of tension structures that consist of
combinations of both tension and compression members, such as tensegrities
and suspended membranes with compression struts. Moreover, it is pointed
out that various functionals can be selected for the purpose of form-finding.

In section 5, some numerical examples of the extended FDM are illustrated
to show that the newly introduced functionals enable us to find the
forms of tension structures that consist of combinations of both tension
and compression members, such as tensegrities and suspended membranes
with compression struts.

In section 6, in which the second half of this work is described,
it is firstly indicated the important role of three expressions used
by the description of the extended FDM, such as stationary problems
of functionals, the principle of virtual work and stationary conditions
using $\nabla$ symbol. They can be commonly found in general problems
of statics, while the original FDM only provides a particular form
of equilibrium equation. Then, to demonstrate the advantage of such
expressions, various form-finding methods are reviewed and compared.
As the result, the common features and the differences over various
form-finding methods can be examined. Finally, to give an overview
of the reviewed methods, the expressions corresponding to them are
shown in the form of three tables.

\section{Force Density Method}

\subsection{Original Formulation}

The FDM is one of the form-finding methods for cable-net structures
which was first proposed by \textit{H. J. Schek} and \textit{K. Linkwitz}
in 1973. When it is explained, two unique points are usually pointed.
The first one is the definition of the force density and the second
one is the linear form of the equilibrium equation provided by the
FDM.

\begin{figure}[!tbh]
\begin{centering}
\includegraphics{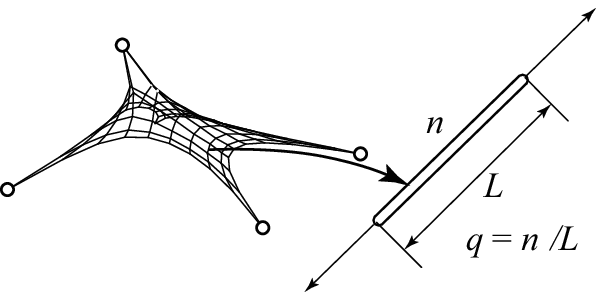}
\par\end{centering}

\begin{centering}
{\footnotesize (a) Definition of force density}\\

\par\end{centering}{\footnotesize \par}

\begin{centering}
\includegraphics{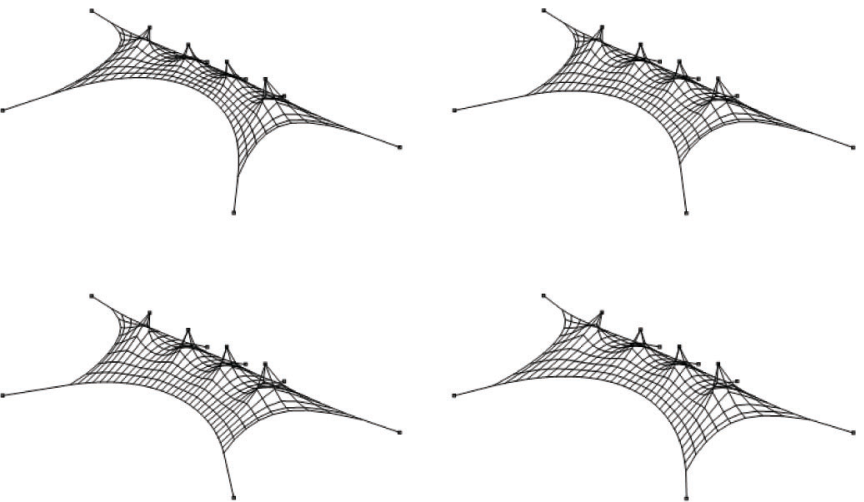}
\par\end{centering}

\begin{centering}
{\footnotesize (b) Form-finding analysis using FDM}
\par\end{centering}{\footnotesize \par}

\caption{\label{fig:FDM}Force Density Method}
\end{figure}

As the first one, the force density $q_{j}$ is defined by 
\begin{equation}
q_{j}=n_{j}/L_{j},\label{eq:DEF_FD}
\end{equation}
where $n_{j}$ and $L_{j}$ denote the tension and length of the $j$-th
member of a structure respectively, as shown in Fig. \ref{fig:FDM}(a).
In the FDM, each tension member is assigned a positive force density
as a prescribed parameter, even though $n_{j}$ and $L_{j}$ are unknown.
However, In Ref. \citep{Schek74}, there is no mention of method to
determine them. Then, it is sometimes pointed out that some trials
must be carried out to obtain an appropriate set of force densities.

As the second one, although the form-finding problems usually formulated
as a non-linear problem, the self-equilibrium equation provided by
the FDM is formulated as a set of simultaneous linear equations. In
detail, when the force densities and the coordinates of the fixed
nodes are prescribed, the self-equilibrium equation of a cable-net
structure is expressed as follows:
\begin{align}
\boldsymbol{D}\cdot\boldsymbol{x} & =-\boldsymbol{D}_{f}\cdot\boldsymbol{x}_{f},\nonumber \\
\boldsymbol{D}\cdot\boldsymbol{y} & =-\boldsymbol{D}_{f}\cdot\boldsymbol{y}_{f},\label{eq:FDM}\\
\boldsymbol{D}\cdot\boldsymbol{z} & =-\boldsymbol{D}_{f}\cdot\boldsymbol{z}_{f},\nonumber 
\end{align}
where $\boldsymbol{D}$ is the equilibrium matrix and $\boldsymbol{x}$,
$\boldsymbol{y}$, and $\boldsymbol{z}$ are the column vectors containing
the coordinates of the nodes. The terms with the subscript $f$ refer
to the fixed nodes, whereas those with no subscript are for the free
nodes.

Using the inverse matrix of $\boldsymbol{D}$, the nodal coordinates
of the free nodes can be simply obtained as follows: 
\begin{align}
\boldsymbol{x} & =-\boldsymbol{D}^{-1}\left(\boldsymbol{D}_{f}\cdot\boldsymbol{x}_{f}\right),\nonumber \\
\boldsymbol{y} & =-\boldsymbol{D}^{-1}\left(\boldsymbol{D}_{f}\cdot\boldsymbol{y}_{f}\right),\label{eq:SFDM}\\
\boldsymbol{z} & =-\boldsymbol{D}^{-1}\left(\boldsymbol{D}_{f}\cdot\boldsymbol{z}_{f}\right),\nonumber 
\end{align}
because, in Eq. \eqref{eq:FDM}, only $\boldsymbol{x}$, $\boldsymbol{y}$,
and $\boldsymbol{z}$ contain the unknown variables.

Because Eq. \eqref{eq:SFDM} simply represents the common procedure
to solve a set of simultaneous linear equations, the FDM can be easily
implemented by general numerical environments. This can be a major
advantage in form-finding analysis of cable-net structures.

Once the nodal coordinates are obtained, the tension in each cable
is calculated by using Eq. \eqref{eq:DEF_FD}. The obtained set of
tension represents a self-equilibrium state of the form, i.e.
\begin{equation}
\boldsymbol{n}=\left\{ q_{1}L_{1},\cdots,q_{m}L_{m}\right\} ,\label{eq:tension-mode}
\end{equation}
where $m$ denotes the number of the members. Generally, such a form
is called a self-equilibrium form and can be used as a prestressed
structure.

Using the FDM, as shown in Fig. \ref{fig:FDM}(b), the form of a cable-net
can be varied by varying the prescribed coordinates of the fixed nodes
and the force densities of the cables.

\subsection{Limitation of FDM}

\begin{figure}[!tbh]
\centering{}\includegraphics{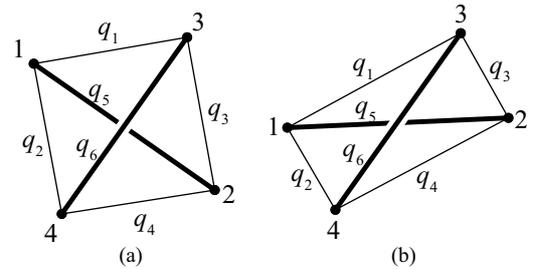}\caption{\label{fig:Self-Equilibrium-Form-of-X-Tensegrity} \textit{X-Tensegrities}}
\end{figure}

In this subsection, the limitation of the FDM is discussed. When it
is applied to self-equilibrium systems that consist of a combination
of both tension and compression members, e.g. tensegrities, some difficulties
arise.

In detail, although it seems possible to assign negative force densities
to the compression members and positive force densities to the tension
members, the FDM can not keep its conciseness any longer as discussed
below.

Let us consider form-finding of a prestressed structure which is called
\textit{X-Tensegrity.} Two different forms of \textit{X-Tensegrity}
are shown by Fig. \ref{fig:Self-Equilibrium-Form-of-X-Tensegrity}
(a) and (b). An \textit{X-Tensegrity} is a planar prestressed structure
that consists of 4 cables (tension) and 2 struts (compression). As
in the case of general tensegrities, the cables connect the struts
and the struts do not touch each other.

For such self-equilibrium systems, due to the absence of the fixed
nodes, Eq. \eqref{eq:FDM} reduces to a simpler form: 
\begin{equation}
\boldsymbol{D}\cdot\boldsymbol{x}=\boldsymbol{0},\,\,\boldsymbol{D}\cdot\boldsymbol{y}=\boldsymbol{0},\,\,\boldsymbol{D}\cdot\boldsymbol{z}=\boldsymbol{0}.\label{eq:FDM_Self-Equilibrium}
\end{equation}

When $\boldsymbol{D}$ is a regular matrix, because it is obvious
that $\boldsymbol{D}^{-1}\cdot\boldsymbol{0}=\boldsymbol{0}$, only
the trivial solution, i.e.
\begin{equation}
\boldsymbol{x}=\boldsymbol{0},\,\,\boldsymbol{y}=\boldsymbol{0},\,\,\boldsymbol{z}=\boldsymbol{0},\label{eq:particular}
\end{equation}
is obtained, which implies that every nodes meet at one point, namely
$[0,0,0]$.

On the other hand, when $\boldsymbol{D}$ is a singular matrix, i.e.
$\mathrm{det}\boldsymbol{D}=0$, Eq. \eqref{fig:Self-Equilibrium-Form-of-X-Tensegrity}
generally has complementary solution, which states the possible forms
of the structure. Such solutions are obtained by analyzing the null
space of $\boldsymbol{D}$. Various methods to analyze the null space
of $\boldsymbol{D}$. Various methods have been proposed to analyze
such a space (see Ref.\citep{Connelly98,Tibert03,Vassart99,zhang06}).
 However, even if the complementary solutions can be obtained by such
methods, the major advantage of the FDM, that the equilibrium equation
can be simply solved by inverse matrix, vanishes.

Let us see a simple example, the form-finding analysis of \textit{X-Tensegrity}
which is shown by Fig. \ref{fig:Self-Equilibrium-Form-of-X-Tensegrity}.
When the FDM is applied to this type of structure, $\boldsymbol{D}$
is calculated by
\begin{align}
\boldsymbol{D} & =\boldsymbol{C}^{T}\boldsymbol{Q}\boldsymbol{C},\\
\boldsymbol{C} & =\left[\begin{array}{cccc}
1 & 0 & -1 & 0\\
1 & 0 & 0 & -1\\
0 & 1 & -1 & 0\\
0 & 1 & 0 & -1\\
1 & -1 & 0 & 0\\
0 & 0 & 1 & -1
\end{array}\right],\\
\boldsymbol{Q} & =\left[\begin{array}{cccccc}
q_{1} & 0 & 0 & 0 & 0 & 0\\
0 & q_{2} & 0 & 0 & 0 & 0\\
0 & 0 & q_{3} & 0 & 0 & 0\\
0 & 0 & 0 & q_{4} & 0 & 0\\
0 & 0 & 0 & 0 & q_{5} & 0\\
0 & 0 & 0 & 0 & 0 & q_{6}
\end{array}\right],
\end{align}
where $\boldsymbol{C}$ is the branch-node matrix (see Ref.\citep{Schek74}
for more detail), $q_{1},\cdots,q_{4}$ are the prescribed force densities
of the cables, and $q_{5},q_{6}$ are of the struts. Then $\boldsymbol{D}$
is represented by:{\scriptsize 
\begin{equation}
\boldsymbol{D}=\left[\begin{array}{cccc}
-q_{1}-q_{2}-q_{5} & q_{5} & q_{1} & q_{2}\\
q_{5} & -q_{3}-q_{4}-q_{5} & q_{3} & q_{4}\\
q_{1} & q_{3} & -q_{1}-q_{3}-q_{6} & q_{6}\\
q_{2} & q_{4} & q_{6} & -q_{2}-q_{4}-q_{6}
\end{array}\right].\label{eq:D}
\end{equation}
}Based on Eq. \eqref{eq:D}, the detail of the form-finding analysis
of \textit{X-Tensegrity} is as follows:
\begin{itemize}
\item \noindent When the assigned force densities, $q_{1},\cdots,q_{6}$,
are in the proportion 1:1:1:1:-1:-1, $\boldsymbol{D}$ becomes a singular
matrix having 3 dimensional null-space. Then, many solutions are obtained.
The components of $\boldsymbol{D}$ and the corresponding complementary
solution are as follows:
\begin{align}
\boldsymbol{D} & =\left[\begin{array}{cccc}
-1 & -1 & 1 & 1\\
-1 & -1 & 1 & 1\\
1 & 1 & -1 & -1\\
1 & 1 & -1 & -1
\end{array}\right],\\
 & (\left\{ q_{1},\cdots,q_{6}\right\} =\left\{ 1,1,1,1,-1,-1\right\} )\\
\boldsymbol{x} & =a\left[\begin{array}{c}
1\\
1\\
1\\
1
\end{array}\right]+b\left[\begin{array}{c}
1\\
-1\\
0\\
0
\end{array}\right]+c\left[\begin{array}{c}
0\\
0\\
1\\
-1
\end{array}\right],\nonumber \\
\boldsymbol{y} & =d\left[\begin{array}{c}
1\\
1\\
1\\
1
\end{array}\right]+e\left[\begin{array}{c}
1\\
-1\\
0\\
0
\end{array}\right]+f\left[\begin{array}{c}
0\\
0\\
1\\
-1
\end{array}\right],\nonumber \\
\boldsymbol{z} & =g\left[\begin{array}{c}
1\\
1\\
1\\
1
\end{array}\right]+h\left[\begin{array}{c}
1\\
-1\\
0\\
0
\end{array}\right]+i\left[\begin{array}{c}
0\\
0\\
1\\
-1
\end{array}\right],\nonumber 
\end{align}
where $a,\cdots,i$ are arbitrary real numbers. This implies, for
example, that both Fig. \ref{fig:Self-Equilibrium-Form-of-X-Tensegrity}(a)
and (b) satisfy Eq. \eqref{eq:FDM_Self-Equilibrium}. The first terms
of the right hand sides denote the position of the center point, namely
$\left[a,d,g\right]$, and the other terms state some symmetries that
all the solutions must have. Note that the particular solution is
just $\boldsymbol{x}=\boldsymbol{y}=\boldsymbol{z}=\boldsymbol{0}$. 
\item \noindent When the assigned force densities, $q_{1},\cdots,q_{6}$,
are not in the proportion 1:1:1:1:-1:-1, $\boldsymbol{D}$ also becomes
a singular matrix but having only 1 dimensional null-space. For example,
if the force densities are in the proportion 2:2:2:2:-1:-1, the components
of $\boldsymbol{D}$ and the corresponding complementary solution
are as follows:
\begin{align}
\boldsymbol{D} & =\left[\begin{array}{cccc}
-3 & -1 & 2 & 2\\
-1 & -3 & 2 & 2\\
2 & 2 & -3 & -1\\
2 & 2 & -1 & -3
\end{array}\right],\\
 & (\left\{ q_{1},\cdots,q_{6}\right\} =\left\{ 2,2,2,2,-1,-1\right\} )\\
\boldsymbol{x} & =a\left[\begin{array}{c}
1\\
1\\
1\\
1
\end{array}\right],\,\,\boldsymbol{y}=b\left[\begin{array}{c}
1\\
1\\
1\\
1
\end{array}\right],\,\,\boldsymbol{z}=c\left[\begin{array}{c}
1\\
1\\
1\\
1
\end{array}\right],\nonumber 
\end{align}
where $a,b,c$ are arbitrary. This implies that all the nodes meet
at one point, namely$\left[a,b,c\right]$.
\end{itemize}

\section{Variational Principle in the FDM}

Let us consider a simple functional

\noindent 
\begin{equation}
\Pi(\boldsymbol{x})=\sum_{j}{w_{j}L_{j}^{2}(\boldsymbol{x})},\label{eq:FDM_Functional}
\end{equation}
where $w_{j}$ and $L_{j}$ denote an assigned positive weight coefficient
and a function to give the length of the $j$-th tension member, respectively.
The column vector $\boldsymbol{x}$ represents unknown variables,
which are $x$, $y$, and $z$ coordinates of the free nodes. It is
generalized as an unknown variable container by
\begin{equation}
\boldsymbol{x}=\left[x_{1}\cdots x_{n}\right]^{T},
\end{equation}
where $n$ denotes the number of the unknown variables. Note that
the coordinates related to the fixed nodes are eliminated from $\boldsymbol{x}$
beforehand and directly substituted in $L_{j}$.

Actually, the FDM can be simply represented by Eq. \eqref{eq:FDM_Functional};
the reason is as follows.

Let $\nabla$ be the gradient operator by
\begin{equation}
\nabla f=\frac{\partial f}{\partial\boldsymbol{x}}\equiv\left[\frac{\partial f}{\partial x_{1}},\cdots,\frac{\partial f}{\partial x_{n}}\right],
\end{equation}
which points the direction of the greatest rate of increase of $f$.
Let $\delta\boldsymbol{x}$ be an arbitrary column vector by
\begin{equation}
\delta\boldsymbol{x}\equiv\left[\begin{array}{c}
\delta x_{1}\\
\vdots\\
\delta x_{n}
\end{array}\right],
\end{equation}
which is called the variation of $\boldsymbol{x}$. Then, the variation
of a function $f(\boldsymbol{x})$ is defined by
\begin{equation}
\delta f(\boldsymbol{x})\equiv\nabla f\cdot\delta\boldsymbol{x}.
\end{equation}

Taking the variation of Eq. \eqref{eq:FDM_Functional}, the stationary
condition of the functional is calculated as follows:
\begin{align}
\delta\Pi=0 & \Leftrightarrow\sum_{j}2w_{j}L_{j}\delta L_{j}=0\\
 & \Leftrightarrow\sum_{j}\left(2w_{j}L_{j}\nabla L_{j}\cdot\delta\boldsymbol{x}\right)=0\\
 & \Leftrightarrow\left(\sum_{j}2w_{j}L_{j}\nabla L_{j}\right)\cdot\delta\boldsymbol{x}=0,
\end{align}
\begin{equation}
\therefore\sum_{j}2w_{j}L_{j}\nabla L_{j}=\boldsymbol{0}.\label{eq:Stat.Cond.FDM}
\end{equation}

\begin{figure}[!tbh]
\begin{centering}
\includegraphics{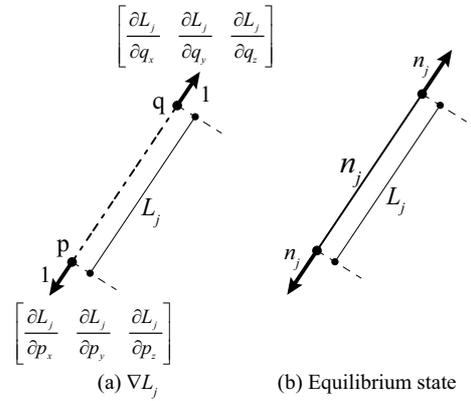}
\par\end{centering}

\caption{\label{fig:Linear-Member}Linear Member}
\end{figure}

In particular case that $\left\{ x_{1},\cdots,x_{n}\right\} $ represents
the Cartesian coordinates of the free nodes, each $L_{j}$ may defined
by the following form:
\begin{align}
 & L_{j}\left(p_{x},\, p_{y},\, p_{z},\, q_{z},\, q_{y},\, q_{z}\right)\nonumber \\
 & \equiv\sqrt{{(p_{x}-q_{x})^{2}+(p_{y}-q_{y})^{2}+(p_{z}-q_{z})^{2}}},\label{eq:Length}
\end{align}
where $\mathrm{p},\mathrm{q}$ denote two ends of $j$-th member and
$p_{x},\cdots,q_{z}$ denote 6 coordinates chosen from $\left\{ x_{1},\cdots,x_{n}\right\} $.
In this case, $\nabla L_{j}$ represents two normalized vectors attached
to both ends of $j$-th member, as shown in Fig. \ref{fig:Linear-Member}(a).

On the other hand, suppose the same member resisting two nodal forces
applied to both ends, as shown in Fig. \ref{fig:Linear-Member}(b).
If the magnitude of the tension of the member is denoted by $n_{j}$,
then the magnitudes of the two nodal forces are also $n_{j}$.

By comparing Fig. \ref{fig:Linear-Member}(a) and (b), a general form
of the self-equilibrium equation for prestressed cable-net structures
is obtained as
\begin{equation}
\sum_{j}{n_{j}\nabla L_{j}}=\boldsymbol{0}.\label{eq:gen_equi}
\end{equation}

To obtain another general form, taking the inner product of Eq. \eqref{eq:gen_equi}
with $\delta\boldsymbol{x}$, the \textbf{Principle of Virtual Work}
for such structures is obtained as
\begin{equation}
\delta w=\sum_{j}{n_{j}\delta L_{j}}=0,\label{eq:gen_pvw}
\end{equation}
where $\delta L_{j}$ is the variation of $L_{j}$.

When a set of $n_{j}$, i.e. 
\begin{equation}
\boldsymbol{n}=\left\{ n_{1},\cdots,n_{m}\right\} ,
\end{equation}
where $m$ denotes the number of the members, satisfies Eq. \eqref{eq:gen_equi},
such a set of $n_{j}$ represents a self-equilibrium state of the
structure.

Remembering the definition of the force density, namely Eq.\eqref{eq:DEF_FD},
Eq. \eqref{eq:gen_equi} can be rewritten as
\begin{equation}
\sum_{j}{q_{j}L_{j}\nabla L_{j}}=\boldsymbol{0},\label{eq:altFDM_Equi}
\end{equation}
which is an alternative form of equilibrium equation provided by the
FDM.

Comparing Eq. \eqref{eq:Stat.Cond.FDM} and Eq. \eqref{eq:altFDM_Equi},
when Eq. \eqref{eq:Stat.Cond.FDM} is considered as a equilibrium
equation, $w_{j}$ is just a half of $q_{j}$. Moreover, when Eq.
\eqref{eq:FDM_Functional} is stationary with a form, it is also the
result obtained by the FDM when the prescribed distribution of $q_{j}$
is as same as $w_{j}$.

Therefore, Eq. \eqref{eq:FDM_Functional}, whose stationary condition
is Eq. \eqref{eq:Stat.Cond.FDM}, is one of the functionals that simply
represents the FDM. In addition, it is assumed that the assigned weight
coefficients would play the same role in form-finding analysis as
the force densities in the FDM.

Because the left hand side of Eq. \eqref{eq:altFDM_Equi} simply represents
the gradient of Eq. \eqref{eq:FDM_Functional}, the stationary problem
of Eq. \eqref{eq:FDM_Functional} can be solve by general direct minimization
approach, such as the steepest decent method or the dynamic relaxation
method \citep{Engeli59,Day65,Papadrakakis81,Papadrakakis82}.

Although Eq. \eqref{eq:FDM} and Eq. \eqref{eq:altFDM_Equi} look
very different, they are accurately identical when each function $L_{j}$
is defined by Eq. \eqref{eq:Length}. Then, let us examine Eq. \eqref{eq:altFDM_Equi}
for further comprehension of the linear form of equilibrium equation
provided by the FDM. If the non-zero components of $\nabla L_{j}$
is split out as 
\begin{align}
\hat{\nabla}L_{j} & \equiv\left[\begin{array}{cccccc}
\frac{\partial L_{j}}{\partial p_{x}} & \frac{\partial L_{j}}{\partial p_{y}} & \frac{\partial L_{j}}{\partial p_{z}} & \frac{\partial L_{j}}{\partial q_{x}} & \frac{\partial L_{j}}{\partial q_{y}} & \frac{\partial L_{j}}{\partial q_{z}}\end{array}\right],
\end{align}
the components of $\hat{\nabla}L_{j}$ are calculated as
\begin{equation}
\hat{\nabla}L_{j}=\left[\begin{array}{cccccc}
\frac{p_{x}-q_{x}}{L_{j}(\boldsymbol{x})} & \frac{p_{y}-q_{y}}{L_{j}(\boldsymbol{x})} & \frac{p_{z}-q_{z}}{L_{j}(\boldsymbol{x})} & \frac{q_{x}-p_{x}}{L_{j}(\boldsymbol{x})} & \frac{q_{y}-p_{y}}{L_{j}(\boldsymbol{x})} & \frac{q_{z}-p_{z}}{L_{j}(\boldsymbol{x})}\end{array}\right].\label{eq:tmp2}
\end{equation}
Here, it can be noticed that $L_{j}(\boldsymbol{x)}$ makes $\hat{\nabla}L_{j}$
non-linear. Then, a linear form can be obtained by multiplying $\hat{\nabla}L_{j}$
with $L_{j}(\boldsymbol{x})$, hence, \textit{
\begin{align}
 & L_{j}\hat{\nabla}L_{j}\nonumber \\
 & =\left[\begin{array}{cccccc}
p_{x}-q_{x} & p_{y}-q_{y} & p_{z}-q_{z} & q_{x}-p_{x} & q_{y}-p_{y} & q_{z}-p_{z}\end{array}\right]\nonumber \\
 & =\left[\begin{array}{cccccc}
1 & 0 & 0 & -1 & 0 & 0\\
0 & 1 & 0 & 0 & -1 & 0\\
0 & 0 & 1 & 0 & 0 & -1\\
-1 & 0 & 0 & 1 & 0 & 0\\
0 & -1 & 0 & 0 & 1 & 0\\
0 & 0 & -1 & 0 & 0 & 1
\end{array}\right]\left[\begin{array}{c}
p_{x}\\
p_{y}\\
p_{z}\\
q_{x}\\
q_{y}\\
q_{z}
\end{array}\right],
\end{align}
}which is the foundation of the linear form of the equilibrium equation
provided by the FDM.

Let us consider a case that each variable $x_{i}$ is also a function
of another set of variables $\{y_{1},\cdots,y_{n}\}$, i.e. 
\begin{align}
x_{1} & =x_{1}(y_{1},\cdots,y_{n})\nonumber \\
 & \vdots\\
x_{n} & =x_{n}(y_{1},\cdots,y_{n}).\nonumber 
\end{align}
In this case, the variation of $\boldsymbol{x}$ is given by

\begin{equation}
\delta\boldsymbol{x}=\boldsymbol{D}\cdot\delta\boldsymbol{y},
\end{equation}
where
\begin{equation}
\boldsymbol{D}\equiv\left[\begin{array}{ccc}
\frac{\partial x^{1}}{\partial y^{1}} & \cdots & \frac{\partial x^{1}}{\partial y^{n}}\\
\vdots & \ddots & \vdots\\
\frac{\partial x^{n}}{\partial y^{1}} & \cdots & \frac{\partial x^{n}}{\partial y^{n}}
\end{array}\right].
\end{equation}
On the other hand, the relation between two types of gradients, namely
with respect to $\boldsymbol{x}$ and $\boldsymbol{y}$, is given
by
\begin{equation}
\left[\begin{array}{ccc}
\frac{\partial f}{\partial y_{1}} & \cdots & \frac{\partial f}{\partial y_{1}}\end{array}\right]=\left[\begin{array}{ccc}
\frac{\partial f}{\partial x_{1}} & \cdots & \frac{\partial f}{\partial x_{1}}\end{array}\right]\cdot\boldsymbol{D}.
\end{equation}
Therefore, 
\begin{align}
\delta f= & \left[\begin{array}{ccc}
\frac{\partial f}{\partial x_{1}} & \cdots & \frac{\partial f}{\partial x_{1}}\end{array}\right]\cdot\delta\boldsymbol{x}\\
= & \left[\begin{array}{ccc}
\frac{\partial f}{\partial x_{1}} & \cdots & \frac{\partial f}{\partial x_{1}}\end{array}\right]\boldsymbol{D}\delta\boldsymbol{y}\\
= & \left[\begin{array}{ccc}
\frac{\partial f}{\partial y_{1}} & \cdots & \frac{\partial f}{\partial y_{1}}\end{array}\right]\cdot\delta\boldsymbol{y},
\end{align}
which implies that the expressions such as Eq. \eqref{eq:Stat.Cond.FDM},
Eq. \eqref{eq:gen_equi}, Eq. \eqref{eq:gen_pvw} and Eq. \eqref{eq:altFDM_Equi}
remain valid when $\{x_{1},\cdots,x_{n}\}$ represents other coordinate,
such as the polar coordinate. 

In this fashion, Eq.\eqref{eq:altFDM_Equi} is the general form of
the equilibrium equation provided by the FDM. On the other hand, the
equilibrium equation provided by the original FDM is one of the particular
forms of the general form, which is only valid for the Cartesian coordinate.

Taking the inner product of Eq. \eqref{eq:altFDM_Equi} with $\delta\boldsymbol{{x}}$,
the \textbf{Principle of Virtual Work} for the FDM is obtained as
\begin{equation}
\delta w=\sum_{j}{q_{j}L_{j}\delta L_{j}}=0.
\end{equation}
Similarly, the \textbf{Principle of Virtual Work} is also deduced
from Eq. \eqref{eq:Stat.Cond.FDM} as: 
\begin{equation}
\delta w=\sum_{j}{2w_{j}L_{j}\delta L_{j}}=0.
\end{equation}
As the result, as well as in the general problems of statics, the
variational principle for the FDM is simply represented by
\begin{equation}
\delta\Pi=0\label{eq:variational principle}
\end{equation}
where $\delta\Pi$ is defined by
\begin{equation}
\delta\Pi\equiv\nabla\Pi\cdot\delta\boldsymbol{x}.
\end{equation}

To conclude this section, it is important to note that, in the original
reference\citep{Schek74}, Eq. \eqref{eq:FDM_Functional} have been
mentioned by the following theorem:
\begin{quotation}
\noindent \textit{''THEOREM 1. Each equilibrium state of an unloaded
network structure with force densities $q_{j}$ is identical with
the net, whose sum of squared way lengths weighted by $q_{j}$ is
minimal. }''
\end{quotation}

\section{Extended Force Density Method }

\subsection{Generalized Formulation of Functional}

\begin{figure}[!tbh]
\centering{}\includegraphics{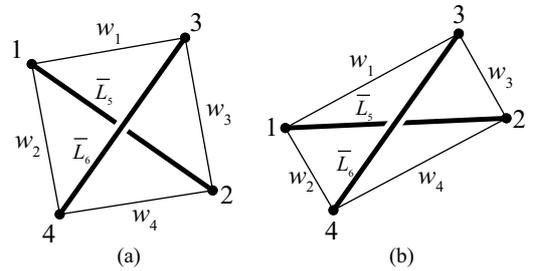}\caption{\label{fig:Self-Equilibrium-Form-of-X-Tensegrity2} \textit{X-Tensegrities
and Prescribed Parameters}}
\end{figure}

\begin{figure*}[!t]
\centering{}\includegraphics{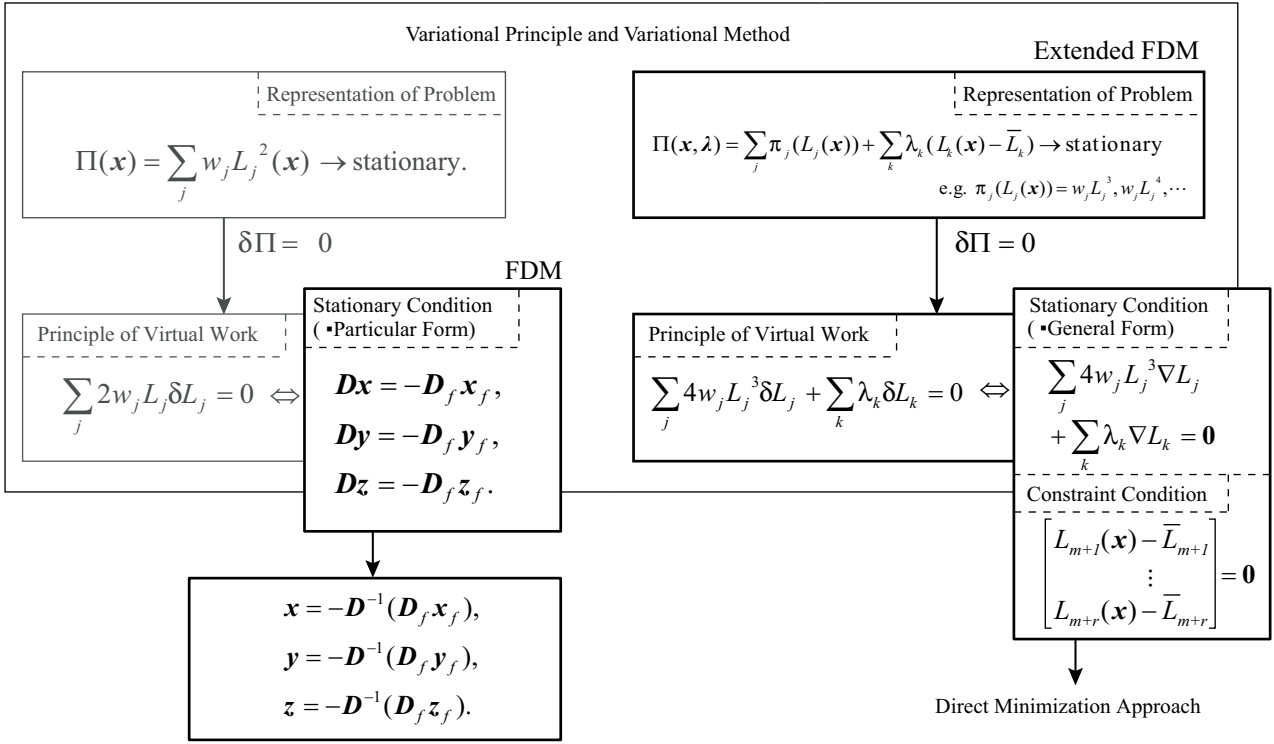}\caption{\label{fig:Diagram}Relation between Original FDM and Extended FDM}
\end{figure*}
In this subsection, the FDM is extended for form-finding of structures
that consist of combinations of both tension and compression members,
e.g. tensegrities.

Let us reconsider the form-finding of \textit{X-Tensegrity} again.
Although it seems possible to assign negative weight coefficients
to the compression members and positive weight coefficients to the
tension members, the same difficulties which is pointed out in subsection
2.2 also arise from the stationary problem of $\Pi(\boldsymbol{x})=\sum w_{j}L_{j}^{2}(\boldsymbol{x})$.
In detail, when the assigned weight coefficients $w_{j}$ are in the
proportion 1:1:1:1:-1:-1, for the 4 tension members and the 2 compression
members respectively, the stationary points forms a space. On the
other hand, when $w_{j}$ are not int the proportion 1:1:1:1:-1:-1,
the stationary points vanish.

First, it is obvious that, without no constraint conditions, every
lengths of the members become simultaneously 0 or infinite. This is
due to the absence of information about the scale of the structure.
Remembering that, in the original FDM, such information is given by
the prescribed coordinates of the fixed nodes, let the lengths of
the compression members be prescribed. Then, using the \textit{Lagrangefs
multiplier method}, a modified functional is obtained as
\begin{equation}
\Pi\left(\boldsymbol{x},\boldsymbol{\lambda}\right)=\sum_{j}{w_{j}L_{j}^{2}\left(\boldsymbol{x}\right)}+\sum_{k}{\lambda_{k}\left(L_{k}\left(\boldsymbol{x}\right)-\bar{{L}}_{k}\right)},\label{eq:mod1}
\end{equation}
where the first sum is taken for all the tension members and the second
is for all the compression members. In addition, $\lambda_{k}$ and
$\bar{L}_{k}$ denote the\textit{ Lagrangefs multiplier} and the prescribed
length of the $k$-th compression member, respectively. Note that
the positive weight coefficients $w_{j}$ are assigned to only the
tension members and the prescribed lengths $\bar{L}_{k}$ are assigned
to only the compression members as shown in Fig.\ref{fig:Self-Equilibrium-Form-of-X-Tensegrity2}.

However, Eq. \eqref{eq:mod1} does not completely eliminate the above
mentioned difficulties. For example, if the assigned weight coefficients
of the tension members $w_{1},\cdots,w_{4}$ are in the proportion
1:1:1:1, and the prescribed lengths of the compression members $\bar{L}_{5},\bar{L}_{6}$
are in the proportion 1:1, both Fig. \ref{fig:Self-Equilibrium-Form-of-X-Tensegrity2}(a)
and (b) satisfy the stationary condition of Eq. \eqref{eq:mod1}.
By using the Pythagorean theorem, i.e. $c^{2}=a^{2}+b^{2}$, it can
be easily verified that the sum of squared lengths of the tension
members takes the same value for both Fig. \ref{fig:Self-Equilibrium-Form-of-X-Tensegrity2}(a)
and (b). Then, it is assumed that such difficulties depend on the
power of $L_{j}$, i.e. 2.

Thus, other functionals, such as
\begin{equation}
\Pi\left(\boldsymbol{x},\boldsymbol{\lambda}\right)=\sum_{j}{w_{j}L_{j}^{4}\left(\boldsymbol{x}\right)}+\sum_{k}{\lambda_{k}\left(L_{k}\left(\boldsymbol{x}\right)-\bar{{L}_{k}}\right)},\label{eq:mod_func}
\end{equation}
are introduced, because it is possible to use other powers of $L_{j}$
instead of 2.

Solving the stationary problem of Eq. \eqref{eq:mod_func}, Fig. \ref{fig:Self-Equilibrium-Form-of-X-Tensegrity2}
(a) becomes the unique solution when the weight coefficients of the
tension members $w_{1},\cdots,w_{4}$ and the prescribed lengths of
the compression members $\bar{L}_{5},\bar{L}_{6}$ are in the proportion
1:1:1:1 and 1:1 respectively. On the other hand, when they are 1:8:8:1
and 1:1, Fig. \ref{fig:Self-Equilibrium-Form-of-X-Tensegrity2} (b)
becomes the unique solution. Actually, Fig. \ref{fig:Self-Equilibrium-Form-of-X-Tensegrity2}(a)
and (b) are the real numerical results obtained by solving such problems.

By the way, let us discuss the following general formulation of functional:
\begin{equation}
\Pi\left(\boldsymbol{x},\boldsymbol{\lambda}\right)=\sum_{j}\pi_{j}\left(L_{j}\left(\boldsymbol{x}\right)\right)+\sum_{k}{\lambda_{k}\left(L_{k}\left(\boldsymbol{x}\right)-\bar{{L}}_{k}\right)}.\label{eq:gen_func}
\end{equation}
The stationary condition of Eq. \eqref{eq:gen_func} with respect
to $\boldsymbol{{x}}$ is as follows:
\begin{equation}
\frac{\partial\Pi}{\partial\boldsymbol{x}}=\sum_{j}{\frac{\partial\pi_{j}\left(L_{j}\left(\boldsymbol{x}\right)\right)}{\partial L_{j}}\nabla L_{j}}+\sum_{k}{\lambda_{k}\nabla L_{k}}=\boldsymbol{0}.\label{eq:Stat.Cond.gen.func}
\end{equation}
Because Eq. \eqref{eq:Stat.Cond.gen.func} has the same form of Eq.
\eqref{eq:gen_equi}, it can be considered as a equilibrium equation.
Then, when Eq. \ref{eq:gen_func} is stationary, the following non-trivial
set of axial forces must satisfy the general form of equilibrium equation:
\begin{equation}
\left\{ n_{1},\cdots,n_{m+r}\right\} =\left\{ \begin{array}{cccccc}
\frac{\partial\pi_{1}}{\partial L_{1}}, & \cdots & ,\frac{\partial\pi_{m}}{\partial L_{m}}, & \lambda_{m+1}, & \cdots & ,\lambda_{m+r}\end{array}\right\} ,\label{eq:tension_mode}
\end{equation}
which represents a self-equilibrium state of structure, where $m$
and $r$ denote the numbers of the tenion and the compression members
respectively.

On the other hand, the stationary condition of Eq. \eqref{eq:gen_func}
with respect to $\boldsymbol{{\lambda}}$ is given by
\begin{equation}
\frac{\partial\Pi}{\partial\boldsymbol{\lambda}}=\left[\begin{array}{c}
\frac{\partial\Pi}{\partial\lambda_{m+1}}\\
\vdots\\
\frac{\partial\Pi}{\partial\lambda_{m+r}}
\end{array}\right]=\left[\begin{array}{c}
L_{m+1}(\boldsymbol{x})-\bar{L}_{m+1}\\
\vdots\\
L_{m+r}(\boldsymbol{x})-\bar{L}_{m+r}
\end{array}\right]=\left[\begin{array}{c}
0\\
\vdots\\
0
\end{array}\right].
\end{equation}

Therefore, any functional that compatible to Eq. \eqref{eq:gen_func}
has a possibility to be used for such form-finding problems. From
now on, let us call $\pi_{j}$ the element functional. Then the following
policy is proposed:
\begin{itemize}
\item Perform form-finding analysis by solving a stationary problem that
is formulated by freely selected element functionals.
\end{itemize}
Taking the inner product of Eq. \eqref{eq:Stat.Cond.gen.func} with
$\delta\boldsymbol{x}$, the \textbf{Principle of Virtual Work} is
obtained as:
\begin{equation}
\delta w=\sum_{j}{\frac{\partial\pi_{j}\left(L_{j}\left(\boldsymbol{x}\right)\right)}{\partial L_{j}}\delta L_{j}}+\sum_{k}{\lambda_{k}\delta L_{k}}=0.\label{eq:pvw_gen_func}
\end{equation}

Additionally, replacing the partial derivatives in Eq. \ref{eq:pvw_gen_func}
by $n_{j}$, the following form can be also used as the \textbf{Principle
of Virtual Work} for general prestressed structures that consist of
combinations of both tension and compression members:
\begin{equation}
\delta w=\sum_{j}{n_{j}(L_{j})\delta L_{j}}+\sum_{k}{\lambda_{k}\delta L_{k}}=0.\label{eq:pvw_gen_func2}
\end{equation}

Comparing Eq. \eqref{eq:pvw_gen_func} and Eq. \eqref{eq:pvw_gen_func2},
if $w_{j}L_{j}^{2}$ is selected as the element functional, the following
relations are derived: 
\begin{equation}
n_{j}=\frac{\partial w_{j}L_{j}^{2}}{\partial L_{j}}=2w_{j}L_{j},\,\therefore w_{j}=n_{j}/2L_{j}.
\end{equation}
Hence, $w_{j}$ can be considered as a half of the force density of
the $j$-th member. 

On the other hand, if $w_{j}L_{j}^{4}$ is selected, then,
\begin{equation}
n_{j}=\frac{\partial w_{j}L_{j}^{4}}{\partial L_{j}}=4w_{j}L_{j}^{3},\,\therefore w_{j}=n_{j}/4L_{j}^{3}.
\end{equation}
Thus, in this fashion, various quantities that are similar to the
force density can be defined. Then, let us call the new quantities,
such as $w_{j}=n_{j}/4L_{j}^{3}$, the extended force density.

Apartting from the linear form of the equilibrium equation, now, the
main characteristics of the original FDM are reconsidered as follows:
\begin{itemize}
\item The coordinates of the fixed nodes are prescribed as constraint conditions.
\item The force densities $q_{j}=n_{j}/L_{j}$ are assigned to each tension
member as known parameters.
\end{itemize}
On the other hand, for example, when $w_{j}L_{j}^{4}$ is selected
as the element functional, the main characteristics of the extended
FDM are as follows:
\begin{itemize}
\item The coordinates of the fixed nodes and the lengths of the compression
members are prescribed as constraint conditions.
\item The extended force densities, e.g. $w_{j}=n_{j}/4L_{j}^{3}$, are
assigned to each tension member as known parameters.
\end{itemize}
Therefore, the extended FDM can be considered as similar method to
the original FDM.

Considering both approach as solving the stationary problems, their
main difference is related to the form of the stationary conditions
and the selection of the computational methods. In the original FDM,
they are as follows:
\begin{itemize}
\item The stationary condition of functional is represented by a particular
form.
\item The stationary condition is simply solved by using an inverse matrix
$\boldsymbol{D}^{-1}$.
\end{itemize}
\noindent On the other hand, in the extended FDM, they are as follows:
\begin{itemize}
\item The stationary condition of functional is represented by a general
form.
\item The stationary condition is solved by general direct minimization
approaches.
\end{itemize}
As an overview of the relation between the original and the extended
FDM, Fig. \ref{fig:Diagram} shows a diagram of both procedures.

\subsection{Additional Analyses}

\noindent 
\begin{figure}[!tbh]
\centering{}\includegraphics{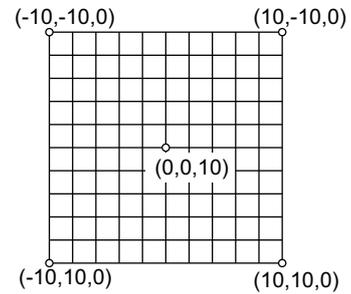}\caption{\label{fig:Analytical-Model-1}Analytical Model}
\end{figure}
\begin{figure*}[!t]
\begin{centering}
\begin{tabular}{cccc}
\includegraphics{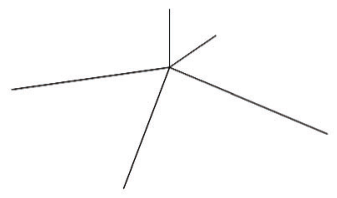} & \includegraphics{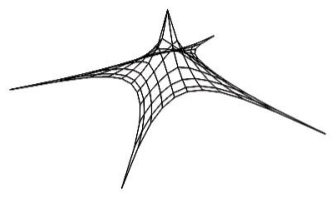} & \includegraphics{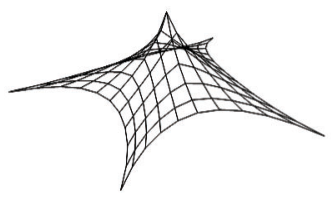} & \includegraphics{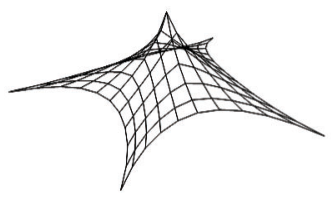}\tabularnewline
{\footnotesize (i) $\sum{L_{j}}\rightarrow\mathrm{min}$} & {\footnotesize (ii) $\sum{L_{j}^{2}}\rightarrow\mathrm{min}$} & {\footnotesize (iii) $\sum{L_{j}^{3}}\rightarrow\mathrm{min}$} & {\footnotesize (iv) $\sum{L_{j}^{4}}\rightarrow\mathrm{min}$}\tabularnewline
\end{tabular}
\par\end{centering}

\caption{\label{fig:Optimization-Results-of-1}Optimization Results of Cable-nets}

\begin{centering}
\begin{tabular}{cccc}
\includegraphics{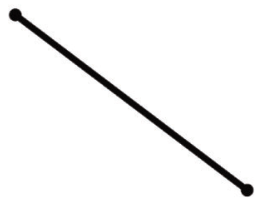} & \includegraphics{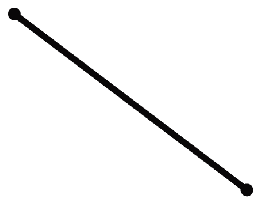} & \includegraphics{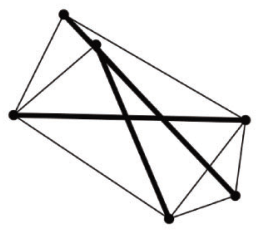} & \includegraphics{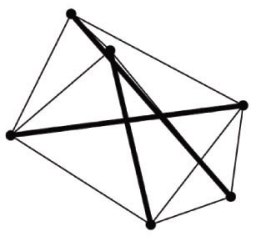}\tabularnewline
{\footnotesize (i) $\sum{L_{j}}\rightarrow\mathrm{min}$} & {\footnotesize (ii) $\sum{L_{j}^{2}}\rightarrow\mathrm{min}$} & {\footnotesize (iii) $\sum{L_{j}^{3}}\rightarrow\mathrm{min}$} & {\footnotesize (iv) $\sum{L_{j}^{4}}\rightarrow\mathrm{min}$}\tabularnewline
\end{tabular}
\par\end{centering}

\caption{\label{fig:Optimization-Results-of-2}Optimization Results of \textit{Simplex
Tensegrities}}
\end{figure*}
In this subsection, some additional numerical analyses are reported
to supplement the concept of the extended FDM. 

Let us consider a net that consists of 220 cables (tension members)
connecting one another and having 5 fixed nodes as shown in Fig. \ref{fig:Analytical-Model-1}.
The prescribed coordinates of the fixed nodes are also shown in the
figure. 

Next, let us find the forms taking minimum numbers of $\sum_{j}L_{j},\cdots,\sum_{j}L_{j}^{4}$,
namely

\begin{equation}
\sum_{j}L_{j}^{\, p}\rightarrow\mathrm{min,}\,(p\in\left\{ 1,2,3,4\right\} ),
\end{equation}
where $L_{j}$ denotes the length of the $j$-th cable. The results
of minimization processes are shown in Fig. \ref{fig:Optimization-Results-of-1}.

On the other hand, Fig. \ref{fig:Optimization-Results-of-2} shows
the other results of the same series of minimization processes performed
on another model, which is based on \textit{Simplex Tensegrity}. A\textit{
Simplex Tensegrity} is a prestressed structure that consists of 9
cables(tension) and 3 struts(compression). In addition, the minimization
processes were only performed on the cables, whereas, the lengths
of the struts were kept constant at prescribed length, 10.0, during
the processes.

Comparing particularly Fig. \ref{fig:Optimization-Results-of-1}(ii)
and Fig. \ref{fig:Optimization-Results-of-2}(ii), $w_{j}L_{j}^{2}$
seems not good for form-finding of tensegrities.

For more detail, when $L_{j}=0$, $\nabla L_{j}$ can no be defined
because $\nabla L$ becomes division by zero (see Eq. \eqref{eq:tmp2}).
Therefore, three of the results, i.e. Fig. \ref{fig:Optimization-Results-of-1}(i),
Fig. \ref{fig:Optimization-Results-of-2}(i) and Fig. \ref{fig:Optimization-Results-of-2}(ii),
are only the solutions of minimization problems, whereas the others
are also the solutions of stationary problems.

\section{Numerical Examples}

In this section, numerical examples of the extended FDM are presented.

In the examples, the stationary problems are represented in the following
form: 
\begin{equation}
\Pi\left(\boldsymbol{x},\boldsymbol{\lambda}\right)=\Pi_{w}(\boldsymbol{x})+\sum_{k}{\lambda_{k}\left(L_{k}\left(\boldsymbol{x}\right)-\bar{{L}}_{k}\right)}.\label{eq:gen_func-1}
\end{equation}
Then, for simplicity, the problems were solved by general direct minimization
approaches, in which just $\Pi_{w}(\boldsymbol{x})$ were minimized
as objective functions and the lengths of the struts were kept constant
at the prescribed lengths $\bar{{L}}_{k}$ during each minimization
process. Hence, only $\boldsymbol{x}$, or the form, was obtained
in each problem.

\subsection{Structures Consisting of Cables and Struts}

\begin{figure}[!tbh]
\centering{}\includegraphics{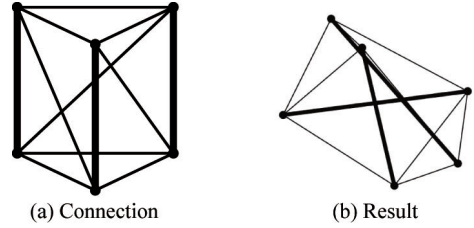}\caption{\label{fig:Simplex-Tensegrity}Simplex Tensegrity}
\end{figure}
\begin{figure}[!tbh]
\centering{}\includegraphics{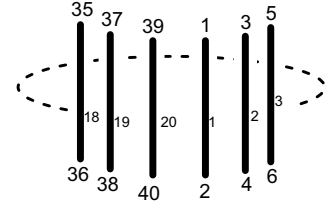}\caption{\label{fig:Instruction-for-Configuration}Sequential Numbers on Struts}
\end{figure}
\begin{table}[!tbh]
\begin{centering}
{\scriptsize }%
\begin{tabular}{|c|c||c|c|}
\hline 
{\scriptsize Cable\#~($w_{1}$)} & {\scriptsize Node\#} & {\scriptsize Cable\#~($w_{2}$)} & {\scriptsize Node\#}\tabularnewline
\hline 
{\scriptsize 1} & {\scriptsize 1-3} & {\scriptsize 41} & {\scriptsize 1-4}\tabularnewline
\hline 
{\scriptsize 2} & {\scriptsize 2-4} & {\scriptsize 42} & {\scriptsize 2-5}\tabularnewline
\hline 
$\vdots$ & {\scriptsize $\vdots$} & $\vdots$ & {\scriptsize $\vdots$}\tabularnewline
\hline 
{\scriptsize 39} & {\scriptsize 39-1} & {\scriptsize 79} & {\scriptsize 39-2}\tabularnewline
\hline 
{\scriptsize 40} & {\scriptsize 40-2} & {\scriptsize 80} & {\scriptsize 40-3}\tabularnewline
\hline 
\multicolumn{4}{c}{{\scriptsize ($C_{1}$)}}\tabularnewline
\end{tabular}{\scriptsize }\\
{\scriptsize }%
\begin{tabular}{|c|c||c|c|}
\hline 
{\scriptsize Cable\#~($w_{1}$)} & {\scriptsize Node\#} & {\scriptsize Cable\#~($w_{2}$)} & {\scriptsize Node\#}\tabularnewline
\hline 
{\scriptsize 1} & {\scriptsize 1-5} & {\scriptsize 41} & {\scriptsize 1-6}\tabularnewline
\hline 
{\scriptsize 2} & {\scriptsize 2-6} & {\scriptsize 42} & {\scriptsize 2-7}\tabularnewline
\hline 
$\vdots$ & {\scriptsize $\vdots$} & $\vdots$ & {\scriptsize $\vdots$}\tabularnewline
\hline 
{\scriptsize 39} & {\scriptsize 39-3} & {\scriptsize 79} & {\scriptsize 39-4}\tabularnewline
\hline 
{\scriptsize 40} & {\scriptsize 40-4} & {\scriptsize 80} & {\scriptsize 40-5}\tabularnewline
\hline 
\multicolumn{4}{c}{{\scriptsize ($C_{2}$)}}\tabularnewline
\end{tabular}{\scriptsize }\\
{\scriptsize $\vdots$}\\
{\scriptsize }%
\begin{tabular}{|c|c||c|c|}
\hline 
{\scriptsize Cable\#~($w_{1}$)} & {\scriptsize Node\#} & {\scriptsize Cable\#~($w_{2}$)} & {\scriptsize Node\#}\tabularnewline
\hline 
{\scriptsize 1} & {\scriptsize 1-19} & {\scriptsize 41} & {\scriptsize 1-20}\tabularnewline
\hline 
{\scriptsize 2} & {\scriptsize 2-20} & {\scriptsize 42} & {\scriptsize 2-21}\tabularnewline
\hline 
$\vdots$ & {\scriptsize $\vdots$} & $\vdots$ & {\scriptsize $\vdots$}\tabularnewline
\hline 
{\scriptsize 39} & {\scriptsize 39-17} & {\scriptsize 79} & {\scriptsize 39-18}\tabularnewline
\hline 
{\scriptsize 40} & {\scriptsize 40-18} & {\scriptsize 80} & {\scriptsize 40-19}\tabularnewline
\hline 
\multicolumn{4}{c}{{\scriptsize ($C_{9}$)}}\tabularnewline
\end{tabular}
\par\end{centering}{\scriptsize \par}

\caption{\label{tab:Connections-by-Cables}Connections by Cables}
\end{table}
As mentioned in section 4.2, a form of the \textit{Simplex Tensegrity}
that consists of 9 cables(tension) and 3 struts(compression) can be
obtained by solving the following problem:\textbf{
\begin{align}
\Pi\left(\boldsymbol{x},\boldsymbol{\lambda}\right)= & \sum_{j}{L_{j}^{4}\left(\boldsymbol{x}\right)}+\sum_{k}{\lambda_{k}\left(L_{k}\left(\boldsymbol{x}\right)-\bar{{L_{k}}}\right)}\nonumber \\
 & \rightarrow\mathrm{{stationary}}.\label{eq:L4-1}
\end{align}
}

Here, in the relation with Eq. \eqref{eq:gen_func-1}, the objective
function $\Pi_{w}$ is $\sum_{j}{L_{j}^{4}\left(\boldsymbol{x}\right)}$.

The \textbf{Principle of Virtual Work} corresponding to Eq. \eqref{eq:L4-1}
is as follows:
\begin{equation}
\delta w=\sum_{j}{4L_{j}^{3}\delta L_{j}}+\sum_{k}{\lambda_{k}\delta L_{k}}=0.
\end{equation}

In the analysis, every prescribed lengths of the struts,$\bar{{L}}_{k}$,
were set to 10.0. The connection between the struts and the cables
in a \textit{Simplex Tensegrity} is as shown in Fig. \ref{fig:Simplex-Tensegrity}
(a). The obtained result is shown by Fig \ref{fig:Simplex-Tensegrity}
(b).

Generally, in the direct minimization approaches (see Ref. \citep{Engeli59,Day65,Papadrakakis81,Papadrakakis82}),
different initial configurations of $\boldsymbol{x}$ may give different
results, because the functionals are basically multimodal.

Then, diffrent random numbers from -2.5 to 2.5 were roughly set to
the initial configuration of \textbf{$\boldsymbol{x}$} in each analysis
in order to obtain local minimums as many as possible, because it
is not only the global minimum but any local minimum has an ability
to be used as a tension structure.

In this example, particularly, only Fig. \ref{fig:Simplex-Tensegrity}
(b) were constantly obtained. However, the same strategy was used
in the following examples and in some of them, many local minimums
were obtained.

Let us consider more complex tensegrities such as a system that consists
of 80 cables (tension) and 20 struts (compression). Let us assign
sequential node numbers to all the ends of the struts, as shown in
Fig. \ref{fig:Instruction-for-Configuration}. 

Even there are a variety of connections between the struts by the
cables, 9 of connections were tested. For each connection, the node
numbers that each cable connects are as shown in Tab. \ref{tab:Connections-by-Cables}.

In this example, the following stationary problem was formulated and
a series of form-finding analyses were carried out:
\begin{align}
\Pi\left(\boldsymbol{x},\boldsymbol{\lambda}\right)= & \sum_{j=1}^{40}{w_{1}L_{j}^{4}\left(\boldsymbol{x}\right)}+\sum_{j=41}^{80}{w_{2}L_{j}^{4}\left(\boldsymbol{x}\right)}+\sum_{k}{\lambda_{k}\left(L_{k}\left(\boldsymbol{x}\right)-\bar{{L_{k}}}\right)}\nonumber \\
 & \rightarrow\mathrm{{stationary}},\label{eq:L4}
\end{align}
in which the cables were divided into two groups and $w_{1}$ denotes
the common weight coefficients for the first group, whereas $w_{2}$
is for the second group. In addition, every prescribed length of the
struts,$\bar{{L}}_{k}$, were constantly set to 10.0.

The \textbf{Principle of Virtual Work} corresponding to Eq. \eqref{eq:L4}
is as follows:
\begin{equation}
\delta w=\sum_{j=1}^{40}{4w_{1}L_{j}^{3}\delta L_{j}}+\sum_{j=41}^{80}{4w_{2}L_{j}^{3}\delta L_{j}}+\sum_{k}{\lambda_{k}\delta L_{k}}=0.
\end{equation}

\begin{figure*}[!t]
\begin{centering}
\includegraphics{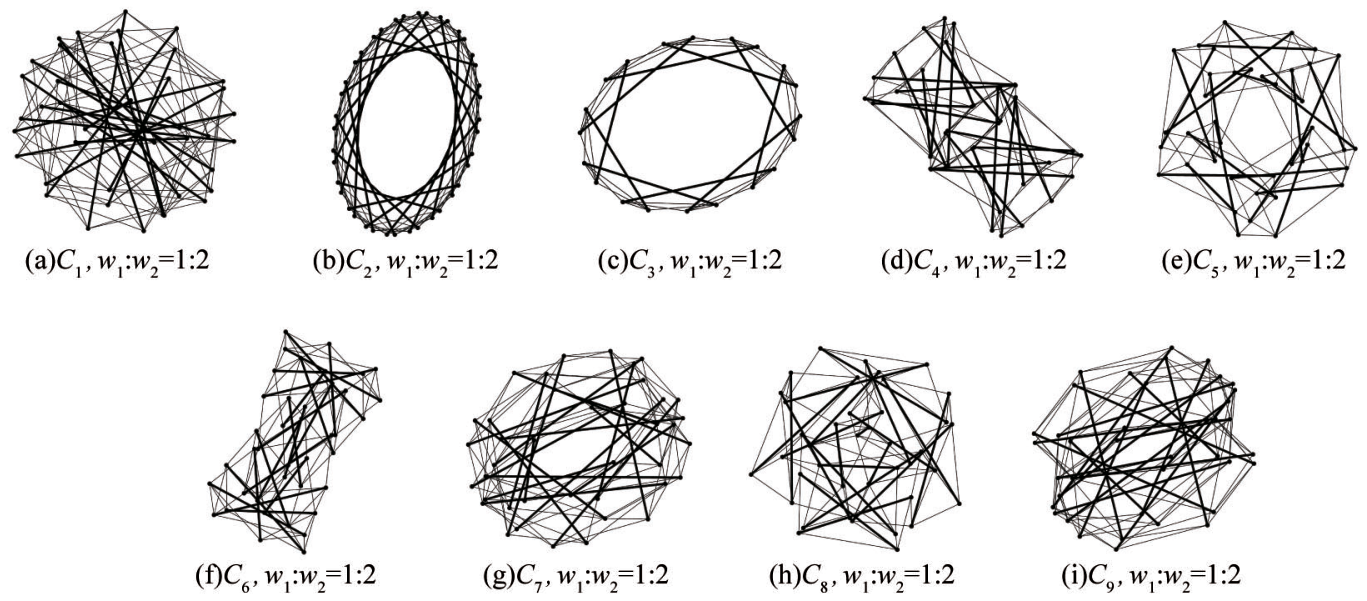}\caption{\label{fig:Discovered-Tensegrities}Discovered Tensegrities}

\par\end{centering}

\begin{centering}
\includegraphics{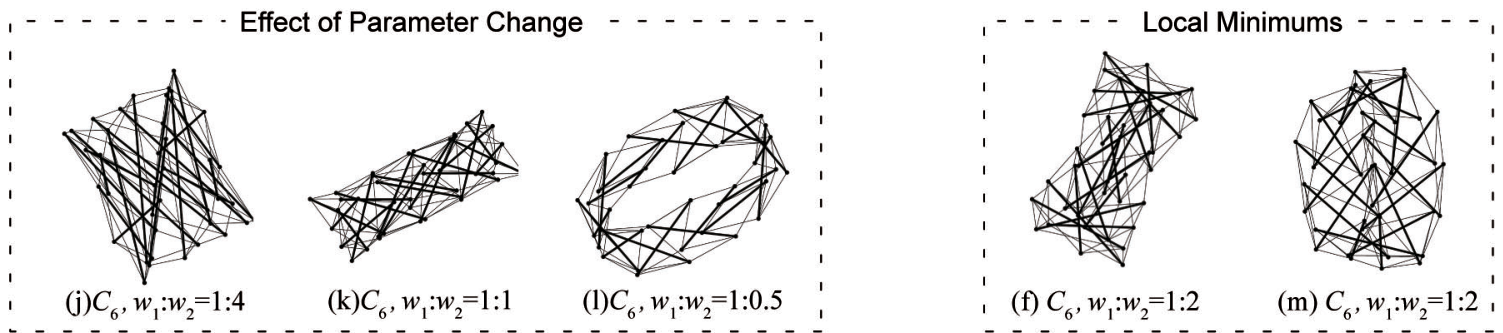}
\par\end{centering}

\caption{\label{fig:Variety-of-Weight}Variety of Forms ($C_{6}$) }
\end{figure*}

When $w_{1}:w_{2}=1:2$, Fig. \ref{fig:Discovered-Tensegrities} shows
the most frequently obtained results for each connection. Fig. \ref{fig:Variety-of-Weight}
(j) to (l) shows how the form varied when the proportion between $w_{1}$
and $w_{2}$ was varied. Interestingly, between Fig. \ref{fig:Variety-of-Weight}
(k) and (l), a transition of the form was observed.

It must be noted that the results shown by Fig. \ref{fig:Discovered-Tensegrities}
are just a fraction of various obtained results and a lot of local
minimums were obtained for each connection, which implies that the
functionals are multimodal. An example of such local minimums are
given by Fig. \ref{fig:Variety-of-Weight} (f) and (m). Although both
results have exactly the same connection and the prescribed parameters,
except the initial configuration of $\boldsymbol{x}$, their forms
look completely different. This is due to the random numbers which
were set to $\boldsymbol{x}$ in each initial step.

\subsection{Structures Consisting of Cables, Membranes and Struts}

\begin{figure}[!tbh]
\centering{}\includegraphics{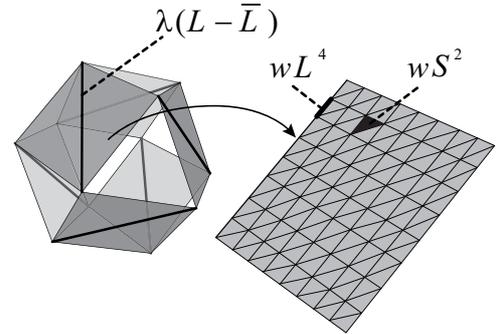}\caption{\label{fig:Analytical-Model}Analytical Model}
\end{figure}
For form-finding of structures that consist of combinations of cables
(tension), membranes (tension), and struts (compression), if the cables
are represented by a set of linear elements and the membranes, by
a set of triangular elements, Eq. \eqref{eq:gen_func} can be extended
as follows:
\begin{align}
\Pi\left(\boldsymbol{x},\boldsymbol{\lambda}\right)= & \sum_{j}{\pi_{j}\left(L_{j}\left(\boldsymbol{x}\right)\right)}+\sum_{k}{\pi{}_{k}\left(S_{k}\left(\boldsymbol{x}\right)\right)}+\nonumber \\
 & \sum_{l}{\lambda_{l}\left(L_{l}\left(\boldsymbol{x}\right)-\bar{L}_{l}\right)},\label{eq:gen-func2}
\end{align}
where the first sum is taken for all the linear elements, the second
is for all the triangular elements and the third is for all the struts.
In addition, $L_{j}$ and $S_{k}$ are defined as the functions to
give the length of the $j$-th linear element and the area of the
$k$-th triangular element respectively. 

The stationary condition of Eq. \eqref{eq:gen-func2} with respect
to $\boldsymbol{x}$ is as follows:

\begin{align*}
\frac{\partial\Pi}{\partial\boldsymbol{x}}= & \nabla\Pi=\sum_{j}{\frac{\partial\pi_{j}\left(L_{j}\left(\boldsymbol{x}\right)\right)}{\partial L_{j}}\nabla L_{j}}+\\
 & \sum_{k}{\frac{\partial\pi{}_{k}\left(S_{k}\left(\boldsymbol{x}\right)\right)}{\partial S_{k}}\nabla S_{k}}+\sum_{l}{\lambda_{l}\nabla L_{l}}=\boldsymbol{0}.
\end{align*}
Replacing the partial differential factors by
\begin{equation}
n_{j}=\frac{\partial\pi_{j}\left(L_{j}\left(\boldsymbol{x}\right)\right)}{\partial L_{j}},\,\,\sigma_{k}=\frac{\partial\pi{}_{k}\left(S_{k}\left(\boldsymbol{x}\right)\right)}{\partial S_{k}},
\end{equation}
a general form that can be considered as a self-equilibrium equation
for such systems is obtained as:
\begin{equation}
\sum_{j}{n_{j}\nabla L_{j}}+\sum_{k}{\sigma_{k}\nabla S_{k}}+\sum_{l}{\lambda_{l}\nabla L_{l}}=\boldsymbol{0}.\label{eq:LSpvw}
\end{equation}
Taking the inner product of Eq. \eqref{eq:LSpvw} with $\delta\boldsymbol{x}$,\emph{
}the \textbf{Principle of Virtual Work} corresponding to Eq. \eqref{eq:LSpvw}
is obtained as follows:

\noindent \textbf{
\begin{equation}
\delta w=\sum_{j}{n_{j}\delta L_{j}}+\sum_{k}{\sigma_{k}\delta S_{k}}+\sum_{l}{\lambda_{l}\delta L_{l}}=0.
\end{equation}
}

In order to alter the cables in the tensegrities by tension membranes,
a form-finding analysis based on the above formulations was carried
out with an analytical model shown by Fig. \ref{fig:Analytical-Model}.
The model is based on the cuboctahedron and consists of 24 cables,
6 membranes, and 6 struts. In detail, every members were translated
to purely geometric components such as curves, surfaces and lines,
then, each curve were discretized by 8 linear elements and each surface
was discretized by 128 triangular elements.

In the analysis, the following stationary problem was formulated and
solved:
\begin{align}
\Pi\left(\boldsymbol{x},\boldsymbol{\lambda}\right)= & \sum_{j}{w_{j}L_{j}^{4}\left(\boldsymbol{x}\right)}+\sum_{k}{w_{k}S_{k}^{2}\left(\boldsymbol{x}\right)}+\nonumber \\
 & \sum_{l}{\lambda_{l}\left(L_{l}\left(\boldsymbol{x}\right)-\bar{L}_{l}\right)}\rightarrow\mathrm{{stationary}}.\label{eq:LSL}
\end{align}
The \textbf{Principle of Virtual Work} corresponding to Eq. \eqref{eq:LSL}
is as follows:
\begin{equation}
\delta w=\sum_{j}{4w_{j}L_{j}^{3}\delta L_{j}}+\sum_{k}{2w_{k}S_{k}\delta S_{k}}+\sum_{l}{\lambda_{l}\delta L_{l}}=0.\label{eq:sss}
\end{equation}

\begin{figure*}[!t]
\begin{centering}
\includegraphics{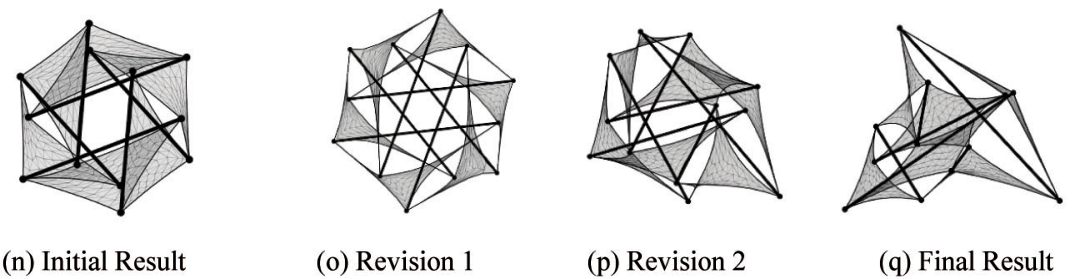}
\par\end{centering}

\caption{\label{fig:Result_of_Complex_Tensegrity}Result of Complex Tensegrity}

\centering{}\includegraphics{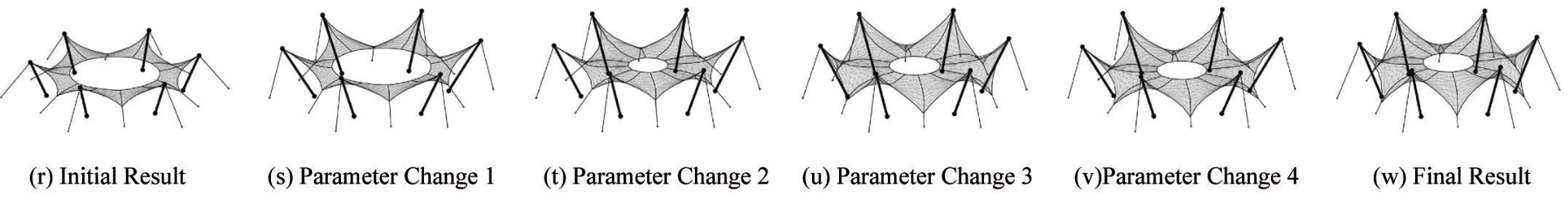}\caption{\label{fig:Form-finding-of-Suspended}Form-finding of Suspended Membrane}
\end{figure*}

At first, all of the weight coefficients of the linear elements were
set to 2.0, those of the triangular elements, 1.0, and the prescribed
lengths of the struts, 10.0. Then the initial result shown by Fig.
\ref{fig:Result_of_Complex_Tensegrity} (n) was obtained. By varying
$w_{j}$, $w_{k}$ and $\bar{{L}}_{l}$, the form was able to be varied
as shown in Fig. \ref{fig:Result_of_Complex_Tensegrity} (o) to (q).

\subsection{Structures Consisting of Cables, Membranes, Struts and Fixed Nodes}

A form-finding analysis of a suspended membrane structure based on
the famous Tanzbrunnen was carried out. It is located in Cologne (K\"{o}ln),
Germany, and was designed by \textit{F. Otto} (1957).

In the analysis, the following problem was formulated and solved:
\textbf{
\begin{align}
\Pi\left(\boldsymbol{x},\boldsymbol{\lambda}\right)= & \sum_{j}{w_{j}L_{j}^{4}\left(\boldsymbol{x}\right)}+\sum_{k}{w_{k}S_{k}^{2}\left(\boldsymbol{x}\right)}+\nonumber \\
 & \sum_{l}{\lambda_{l}\left(L_{l}\left(\boldsymbol{x}\right)-\bar{L}_{l}\right)}\rightarrow\mathrm{{stationary}},
\end{align}
}where, as well as in the previous example, the first sum is taken
for all the linear elements, the second is for all the triangular
elements, and the third is for all the struts. As well as in section
3, the prescribed coordinates of the fixed nodes are eliminated from
$\boldsymbol{x}$ beforehand and directly substituted in $L_{j}$
and $S_{k}$.

By varying $w_{j}$, $w_{k}$ and $\bar{{L}}_{l}$, as shown in Fig.
\ref{fig:Form-finding-of-Suspended}, the form was able to be varied.
Note that Fig. \ref{fig:Form-finding-of-Suspended}(w) looks having
a close form to the real one.

\section{Review of Various Form-Finding Methods}

In the description of the extended FDM, which is just introduced in
the previous sections, three diffrent types of expressions are mainly
used, they are, stationary problems of functionals, the principle
of virtual work, and stationary conditions using $\nabla$ symbol.
Such expressions can be commonly found in general problems of statics.

In this section, by using such expressions, various form-finding methods
are reviewed and compared, in the relation with the extended FDM.
The methods to be reviewed are, the original FDM, the surface stress
density method (SSDM) \citep{Maurin98}, and the methods to solve
the minimal surface problem, a variational method for tensegrities
\citep{kazuma06}.

First, let us review the SSDM, which is also an extension of the FDM
and a form-finding method for membrane structures. It was proposed
by B. Maurin et al., in 1998.

In the SSDM, the membranes are discretized by many triangular membrane
elements and in each elements, the Cauchy stress tensor $\sigma_{\cdot\beta}^{\alpha}$
is assumed as uniform and isotropic, i.e. $\sigma_{\cdot\beta}^{\alpha}=\hat{\sigma}\delta_{\cdot\beta}^{\alpha}$,
in order to obtain uniform stress surfaces. As an analogy of the definition
of the force density, the surface stress density $Q_{j}$ in each
element $j$ is defined by 
\begin{equation}
Q_{j}=\sigma_{j}/S_{j},
\end{equation}
where $\sigma_{j}$ is just the scalar multiple of $\hat{\sigma}_{j}$
with the element thickness $t_{j}$ and $S_{j}$ denotes each element
area. Then, an equilibrium equation is formulated by considering the
equilibrium of all nodes of the triangular elements.

\begin{figure}[!tbh]
\begin{centering}
\includegraphics{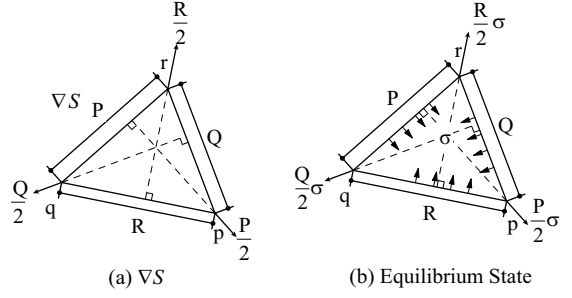}
\par\end{centering}

\caption{\label{fig:Triangular-Element}Triangular Element}
\end{figure}

Let us rewrite the equilibrium equation provided by the SSDM by using
$\nabla$ symbol, which is the same fashion that applied to the original
FDM (see section 3). First, let $S(\boldsymbol{x})$ be a function
to give the area of a triangle determined by three nodes whose 9 coordinates
are included in $\boldsymbol{x}=\left[\begin{array}{ccc}
x_{1} & \cdots & x_{n}\end{array}\right]^{T}$ . When $\nabla S$ is defined by
\begin{equation}
\nabla S\equiv\left[\frac{\partial S}{\partial x_{1}},\cdots,\frac{\partial S}{\partial x_{n}}\right],
\end{equation}
it represents three vectors attached to each node, as shown in Fig.
\ref{fig:Triangular-Element}(a).

By the way, let a triangular membrane element, of which the thickness
is assumed as uniform and denoted by $t$, be resisting three nodal
forces applied to each node. For the Cauchy stress filed in each element,
in the same fashion of the SSDM, let $\sigma_{\cdot\beta}^{\alpha}=\hat{\sigma}\delta_{\cdot\beta}^{\alpha}$
and $\sigma=\hat{\sigma}t$. When such an element is in equilibrium
with the three nodal forces, the nodal forces can be calculated uniquely,
and it is as shown in Fig. \ref{fig:Triangular-Element} (b).

Comparing Fig. \ref{fig:Triangular-Element} (a) and (b), a general
form of self-equilibrium equation for general systems that consist
of such elements is obtained as 
\begin{equation}
\sum_{j}\sigma_{j}\nabla S_{j}=\boldsymbol{0},\label{eq:Gen_Eq_Surf}
\end{equation}
taking the inner product of Eq. \eqref{eq:Gen_Eq_Surf} with $\delta\boldsymbol{x}$,
the \textbf{Principle of Virtual Work} for such a system is obtained
as:
\begin{equation}
\delta w=\sum_{j}{\sigma_{j}\delta S_{j}}=0.
\end{equation}

By the way, in the SSDM, the surface stress density $Q_{j}$ is defined
by
\begin{equation}
Q_{j}=\sigma_{j}/S_{j},\label{eq:DEF_SSD}
\end{equation}
then substituting Eq. \eqref{eq:DEF_SSD} to Eq. \eqref{eq:Gen_Eq_Surf},
a general form for the self equilibrium equation of the SSDM is obtained
as:
\begin{equation}
\sum_{j}{Q_{j}S_{j}\nabla S_{j}}=\boldsymbol{0}.\label{eq:stat_cond_SSDM}
\end{equation}

Then, one of the functionals that simply represents the SSDM is as
follows:

\begin{equation}
\Pi\left(\boldsymbol{x}\right)=\sum_{j}w_{j}S_{j}^{2}\left(\boldsymbol{x}\right),\label{eq:SSDM_Func}
\end{equation}
because the stationary condition of Eq. \eqref{eq:SSDM_Func} is given
by
\begin{equation}
\frac{\partial\Pi(\boldsymbol{x})}{\partial\boldsymbol{x}}=\nabla\Pi=\sum_{j}{2w_{j}S_{j}\nabla S_{j}}=\boldsymbol{0},\label{eq:stat.cond.SSDM}
\end{equation}
and when Eq. \eqref{eq:stat.cond.SSDM} is considered as one of the
equilibrium equations given by Eq. \eqref{eq:Gen_Eq_Surf}, $w_{j}$
can be considered as just a half of $Q_{j}$. In addition, each $w_{j}$
also represents an extended force density such as $w_{j}=\sigma_{j}/2S_{j}$.

Based on the proposed functionals for the original FDM and the SSDM,
namely $\sum_{j}{w_{j}L_{j}^{2}}$ and $\sum_{j}{w_{j}S_{j}^{2}}$,
the SSDM looks truly an extension of the original FDM.

Moreover, based on the corresponding \textbf{Principle of Virtual
Works}, i.e.
\begin{align}
\delta w & =\sum_{j}{2w_{j}L_{j}\delta L_{j}}=0,\label{eq:pvw_FDM}\\
\delta w & =\sum_{j}{2w_{j}S_{j}\delta S_{j}}=0,
\end{align}
$2w_{j}L_{j}$ and $2w_{j}S_{j}$ can be considered as general forces
which act within the members or the elements and tend to produce small
change of $L_{j}$ and $S_{j}$, respectively.

In addition, if the \textbf{Principle of Virtual Works }are written
in the following forms: 
\begin{align}
\delta w & =\sum_{j}{w_{j}\delta\left(L_{j}^{2}\right)}=0,\\
\delta w & =\sum_{j}{w_{j}\delta\left(S_{j}^{2}\right)}=0,
\end{align}
then, the extended force densities, $w_{j}=n_{j}/2L_{j}$ and $w_{j}=\sigma_{j}/2S_{j}$,
can be considered as general forces which act within the members or
the elements and tend to produce small change of $L_{j}^{2}$ and
$S_{j}^{2}$. 

Next, let us compare the following two problems:

\begin{align}
\Pi(\boldsymbol{x})=\sum_{j}{S_{j}(\boldsymbol{x})} & \rightarrow\mathrm{stationary},\label{eq:S1}\\
\Pi(\boldsymbol{x})=\sum_{j}{S_{j}^{\,2}(\boldsymbol{x})} & \rightarrow\mathrm{stationary},\label{eq:S2}
\end{align}
because, for the minimal surface problem, $\sum_{j}{S_{j}}$ is often
used, whereas, $\sum_{j}{S_{j}^{2}}$ simply represents the SSDM when
the distribution of the surface stress densities is given as uniform.

\begin{figure}[!tbh]
\centering{}\includegraphics{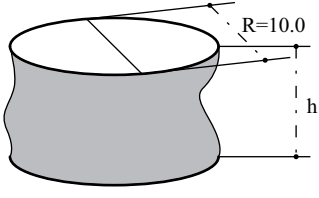}\caption{\label{fig:Instructtion-for-Soap}Form-finding Problem of Simple Membrane}
\end{figure}

\begin{figure}[!tbh]
\begin{centering}
\begin{tabular}{cc}
\includegraphics{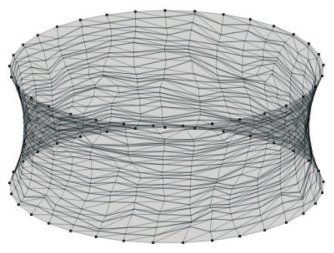} & \includegraphics{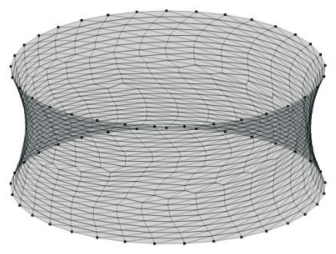}\tabularnewline
{\footnotesize (a-1)$\sum{S}$ h:4.0 area:122} & {\footnotesize (a-2)$\sum{S^{2}}$ h:4.0 area:122}\tabularnewline
\includegraphics{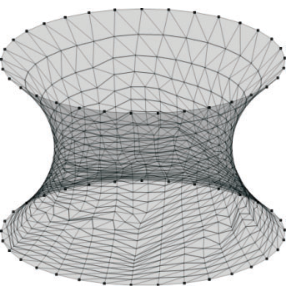} & \includegraphics{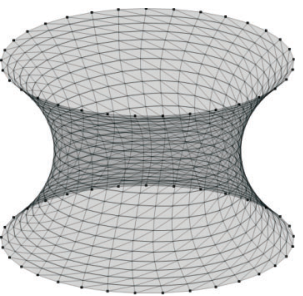}\tabularnewline[5pt]
{\footnotesize (b-1)$\sum{S}$ h:6.5 area:186} & {\footnotesize (b-2)$\sum{S^{2}}$ h:6.5 area:186}\tabularnewline
\end{tabular}
\par\end{centering}

\centering{}\caption{\label{fig:Comparison-of-}Comparison of $\sum S$ and $\sum S^{2}$}
\end{figure}

By applying both problems to the same numerical model shown in Fig.
\ref{fig:Instructtion-for-Soap}, 2 pairs of results were obtained
as shown in Fig. \ref{fig:Comparison-of-}. In addition, such forms
are easily observed by a soap-film experiment.

First of all, due to the fact that they are different functionals,
it is not obvious that the stationary points given by Eq. \eqref{eq:S2}
are minimal surfaces. However, the forms of (a-1) and (b-1) look identical
with (a-2) and (b-2). On the other hand, their mesh distributions
look dissimilar, i.e. the results given by $\sum_{j}{S_{j}^{2}}$
seem to have better mesh distributions in comparison with those by
$\sum_{j}{S_{j}}$. 

\begin{figure}[!tbh]
\centering{}\includegraphics{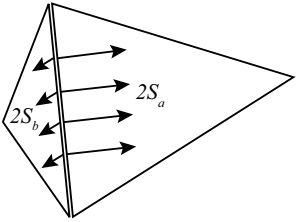}\caption{\label{fig:Stress-State-in}Stress State in SSDM}
\end{figure}

Then, let us see the \textbf{Principle of Virtual Works}, i.e.
\begin{align}
\delta w & =\sum_{j}{\delta S_{j}}=0,\label{eq:pvw6}\\
\delta w & =\sum_{j}{2S_{j}\delta S_{j}}=0.\label{eq:pvw5}
\end{align}
Then, it can be noticed that, in Eq. \eqref{eq:pvw5}, the general
forces which tend to produce small change of $S_{j}$ are proportional
to $S_{j}$, which implies that each element is hard to have bigger
or smaller area compared to the surrounding elements (see Fig. \ref{fig:Stress-State-in}).
On the other hand, in Eq. \eqref{eq:pvw6}, whatever element area
that each element has, the coefficients of $\delta S_{j}$ remain
always 1. Therefore, as long as the total element area is minimum,
each element is able to have bigger or smaller area compared to the
surrounding elements. Thus, the difference appeared in Fig. \ref{fig:Comparison-of-}
can be well explained by the principle of virtual works. 

\begin{figure}[!tbh]
\centering{}\includegraphics{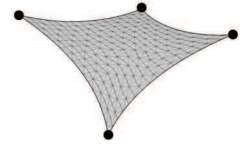}\caption{\label{fig:Membrane-with-Cables}Membrane with Cables}
\end{figure}

The SSDM has been proposed for structures that consist of combinations
of membranes and cables. When the SSDM is applied to such structures,
as same as in section 5.2, the cables are represented by linear elements
and the membranes are represented by triangular elements. Then, the
force densities are assigned to the linear elements and the surface
stress densities are assigned to the triangular elements. In such
cases, the SSDM can be simply represented by

\begin{equation}
\Pi(\boldsymbol{x})=\sum_{j}{w_{j}L_{j}^{2}(\boldsymbol{x})}+\sum_{k}{w_{k}S_{k}^{2}(\boldsymbol{x})}\rightarrow\mathrm{stationary,}\label{eq:SSDM_CABLE_MEMBRANE}
\end{equation}
where the first sum is taken for all the linear elements and the second
is for all the triangular elements. Fig. \ref{fig:Membrane-with-Cables}
shows one of the results given by solving Eq. \eqref{eq:SSDM_CABLE_MEMBRANE}.
The corresponding \textbf{Principle of Virtual Work} is as follows:
\begin{equation}
\delta w=\sum_{j}2w_{j}L_{j}\delta L_{j}+\sum_{k}2w_{k}S_{k}\delta S_{k}=0,\label{eq:pvw7}
\end{equation}
and the stationary condition is obtained as:
\begin{equation}
\frac{\partial\Pi}{\partial\boldsymbol{x}}=\sum_{j}2w_{j}L_{j}\nabla L_{j}+\sum_{k}2w_{k}S_{k}\nabla S_{k}=\boldsymbol{0}.\label{eq:fff}
\end{equation}
Eq. \eqref{eq:SSDM_CABLE_MEMBRANE}-\eqref{eq:fff} are just simple
compositions of corresponding expressions related to the original
FDM and the SSDM, which imply the potential ability of such expressions
for extension.

Next, let us review form-finding methods which have been proposed
to determin the forms of tensegrities. Particularly, let us examine
the following two problems: 
\begin{align}
\Pi(\boldsymbol{x},\boldsymbol{\lambda})= & \sum_{j}{\frac{1}{2}k_{j}\left(L_{j}(\boldsymbol{x)}-\bar{L}_{j}\right)^{2}}\nonumber \\
+ & \sum_{k}{\lambda_{k}(L_{k}(\boldsymbol{x)}-\bar{L}_{k})}\rightarrow\mathrm{stationary},\label{eq:var_tensegrity}
\end{align}
\begin{align}
\Pi(\boldsymbol{x},\boldsymbol{\lambda})= & \sum_{j}{w_{j}L_{j}^{4}(\boldsymbol{x)}}\nonumber \\
+ & \sum_{k}{\lambda_{k}(L_{k}(\boldsymbol{x)}-\bar{L}_{k})}\rightarrow\mathrm{stationary},\label{eq:ext_FDM}
\end{align}
where the first sum is taken for all the cables and the second is
for all the struts.

\begin{figure}[!tbh]
\centering{}%
\begin{tabular}{cc}
\includegraphics{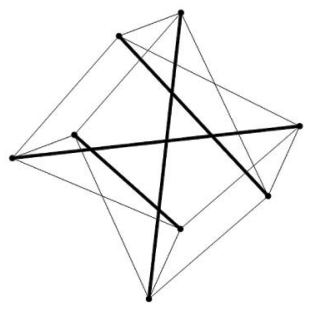} & \includegraphics{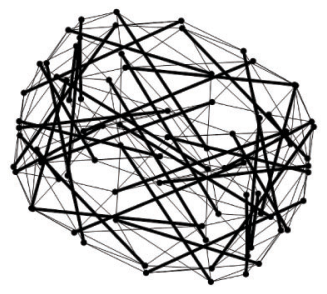}\tabularnewline
{\footnotesize (a) Result by Eq. \eqref{eq:var_tensegrity}} & {\footnotesize (b) Result by Eq. \eqref{eq:ext_FDM}}\tabularnewline
\end{tabular}\caption{\label{fig:Form-Finding-of-Tensegrities}Form-Finding of Tensegrities}
\end{figure}

In Ref. \citep{kazuma06}, Eq. \eqref{eq:var_tensegrity} is proposed
for the form-finding of tensegrities. In Eq. \eqref{eq:var_tensegrity},
$k_{j}$ and $\bar{L}_{j}$ represent virtual stiffness and virtual
initial length of the $j$-th cable respectively, which do not represent
real material but define special (soft) material for form-finding
analysis. Therefore, as discussed below, an appropriate set of $\bar{L}_{j}$
is needed. On the other hand, $\bar{L}_{k}$ represents just the objective
length of the $k$-th strut. Fig. \ref{fig:Form-Finding-of-Tensegrities}(a)
shows an example of tensegrities which was obtained by solving Eq.
\eqref{eq:var_tensegrity} by the authors.

On the other hand, Eq. \eqref{eq:ext_FDM} is one of the stationary
problems which was just proposed in this work. Fig. \ref{fig:Form-Finding-of-Tensegrities}(b)
shows an example of tensegrities given by solving Eq. \eqref{eq:ext_FDM}.

With respect to the the second sums for the struts, there look no
difference.

On the other hand, with respect to the first sums, which are for the
cables, some differences can be recognized. They are, the powers and
the terms that are powered. In addition, while the first sum of Eq.
\eqref{eq:var_tensegrity} looks an analogy of elastic energy of Hook's
spring, the first sum of Eq. \eqref{eq:ext_FDM} looks different.

Then, let us see the \textbf{Principle of Virtual Works} corresponding
to Eq. \eqref{eq:var_tensegrity} and Eq. \eqref{eq:ext_FDM}, i.e.

\begin{equation}
\delta w=\sum_{j}{k_{j}\left(L_{j}-\bar{{L}}_{j}\right)\delta L_{j}}+\sum_{k}{\lambda_{k}\delta L_{k}}=0,\label{eq:pvw8}
\end{equation}
\begin{equation}
\delta w=\sum{4w_{j}L_{j}^{3}\delta L_{j}}+\sum_{k}{\lambda_{k}\delta L_{k}}=0.\label{eq:pvw9}
\end{equation}
Thus, it can be noticed that, in Eq. \eqref{eq:pvw8}, the general
forces $k_{j}\left(L_{j}-\bar{{L}}_{j}\right)$ which tend to produce
small change of $L_{j}$ are proportional to $(L_{j}-\bar{{L}}_{j})$.
Due to the fact that $\left(L_{j}-\bar{{L}}_{j}\right)$ can take
negative numbers, some of the cables may become compression. Then,
it can be noticed that an appropriate set of $\bar{{L}}_{j}$ is needed
to ensure every $(L_{j}-\bar{{L}}_{j})$ be positive.

For this purpose, one of the simplest ideas to determine each $\bar{{L}}_{j}$
in Eq. \eqref{eq:var_tensegrity} for the cables is to set every $\bar{{L}}_{j}$
as 0. However, when every $\bar{{L}}_{j}$ are set to 0 in Eq. \eqref{eq:var_tensegrity}
or Eq. \eqref{eq:pvw8}, some difficulties arise as mentioned in section
4. Then, to eliminate the difficulties, one of the simplest ideas
is to alter the power of the term $\left(L_{j}-\bar{{L}}_{j}\right)$
to other numbers such as 4. Thus, the equations used in the extended
FDM, such as Eq. \eqref{eq:ext_FDM} and Eq. \eqref{eq:pvw9}, emerge.

In addition, the \textbf{Principle of Virtual Work} corresponding
to Eq. \eqref{eq:ext_FDM} is also represented in the following form:
\begin{equation}
\delta w=\sum_{j}{w_{j}\delta\left(L_{j}^{4}\right)}+\sum_{k}{\lambda_{k}\delta L_{k}}=0,
\end{equation}
which states that the extended force densities, i.e. $w_{j}=n_{j}/4L_{j}^{3}$,
can be considered as general forces which act within the cables and
tend to produce small change of $L_{j}^{4}$.

As a result of above discussion, a common feature which is shared
by many form-finding methods have been found. By seeing the following
above mentioned \textbf{Principle of Virtual Works},
\begin{equation}
\delta w=\sum_{j}2w_{j}L_{j}\delta L_{j}=0,
\end{equation}
\begin{equation}
\delta w=\sum_{j}2w_{j}S_{j}\delta S_{j}=0,
\end{equation}
\begin{equation}
\delta w=\sum_{j}\delta S_{j}=0,
\end{equation}
it can be noticed that the general forces which act within the elements
or the members remain always positive.

Finally, the stationary problems of functionals, the principle of
virtual works and the stationary conditions using $\nabla$ symbol,
which were just compared in this section, are shown in Tab. \ref{tab:Table-of-Functionals}
to \ref{tab:Table-of-Stationary} as an overview. By using those three
expressions that are usually found in various problems of statics,
the common features and the differences over various form-finding
methods can be examined, as discussed in this section. Moreover, they
also enable us to combine or extend the methods in natural ways.

\begin{table*}[!t]
\begin{centering}
\caption{\label{tab:Table-of-Functionals}Table of Stationary Problems}
\begin{tabular}{cc|cc}
\hline 
 & \textbf{\scriptsize Force Density Method}{\scriptsize \citep{Schek74}} &  & \textbf{\scriptsize Minimal Surface}{\scriptsize{} e.g. \citep{Bletzinger01}}\tabularnewline
\hline 
 & {\scriptsize cables} &  & {\scriptsize membrane}\tabularnewline
\hline 
{\scriptsize (a)} & {\scriptsize $\Pi=\sum{wL^{2}}\rightarrow\mathrm{stationary.}$} & {\scriptsize (b)} & {\scriptsize $\Pi=\sum{S}\rightarrow\mathrm{stationary.}$}\tabularnewline
\hline 
\hline 
 & \multicolumn{3}{c}{\textbf{\scriptsize Surface stress density method}{\scriptsize \citep{Maurin98}}}\tabularnewline
\hline 
 & {\scriptsize membrane} &  & {\scriptsize membrane with cables}\tabularnewline
\hline 
{\scriptsize (c)} & {\scriptsize $\Pi=\sum{wS^{2}}\rightarrow\mathrm{stationary.}$} & {\scriptsize (d)} & {\scriptsize $\Pi=\sum{wL^{2}}+\sum{wS^{2}}\rightarrow\mathrm{stationary.}$}\tabularnewline
\hline 
\hline 
 & \textbf{\scriptsize Variational Method for Tensegrities}{\scriptsize \citep{kazuma06}} &  & \textbf{\scriptsize Extended Force Density Method}{\scriptsize{} (Proposed)}\tabularnewline
\hline 
 & {\scriptsize cables and struts} &  & {\scriptsize cables and struts}\tabularnewline
\hline 
{\scriptsize (e)} & {\scriptsize $\Pi=\sum{\frac{1}{2}k(L-\bar{L})^{2}}+\sum{\lambda(L-\bar{{L}})}\rightarrow\mathrm{stationary.}$} & {\scriptsize (f)} & {\scriptsize $\Pi=\sum{wL^{4}}+\sum{\lambda(L-\bar{{L}})}\rightarrow\mathrm{stationary.}$}\tabularnewline
\hline 
\end{tabular}
\par\end{centering}

\begin{centering}
\vspace*{\bigskipamount}
\caption{\label{tab:Table-of-Principle}Table of Principle of Virtual Works}

\par\end{centering}

\begin{centering}
\begin{tabular}{cc|cc}
\hline 
 & \textbf{\scriptsize Force Density Method}{\scriptsize \citep{Schek74}} &  & \textbf{\scriptsize Minimal Surface}{\scriptsize{} e.g. \citep{Bletzinger01}}\tabularnewline
\hline 
 & {\scriptsize cables} &  & {\scriptsize membrane}\tabularnewline
\hline 
{\scriptsize (a)} & {\scriptsize $\delta w=\sum{2wL\delta L}=0$} & {\scriptsize (b)} & {\scriptsize $\delta w=\sum{\delta S}=0$}\tabularnewline
\hline 
\hline 
 & \multicolumn{3}{c}{\textbf{\scriptsize Surface stress density method}{\scriptsize \citep{Maurin98}}}\tabularnewline
\hline 
 & {\scriptsize membrane} &  & {\scriptsize membrane with cables}\tabularnewline
\hline 
{\scriptsize (c)} & {\scriptsize $\delta w=\sum{2wS\delta S}=0$} & {\scriptsize (d)} & {\scriptsize $\delta w=\sum{2wL\delta L}+\sum{2wS\delta S}=0$}\tabularnewline
\hline 
\hline 
 & \textbf{\scriptsize Variational Method for Tensegrities}{\scriptsize \citep{kazuma06}} &  & \textbf{\scriptsize Extended Force Density Method}{\scriptsize{} (Proposed)}\tabularnewline
\hline 
 & {\scriptsize cables and struts} &  & {\scriptsize cables and struts}\tabularnewline
\hline 
{\scriptsize (e)} & {\scriptsize $\delta w=\sum{kL\delta L}+\sum{\lambda\delta L}=0$} & {\scriptsize (f)} & {\scriptsize $\delta w=\sum{4wL^{3}\delta L}+\sum{\lambda\delta L}=0$}\tabularnewline
\hline 
\end{tabular}
\par\end{centering}

\begin{centering}
\vspace*{\bigskipamount}

\par\end{centering}

\begin{centering}
\caption{\label{tab:Table-of-Stationary}Table of Stationary Conditions}

\par\end{centering}

\centering{}{\small }%
\begin{tabular}{cc|cc}
\hline 
 & \textbf{\scriptsize Force Density Method}{\scriptsize \citep{Schek74}} &  & \textbf{\scriptsize Minimal Surface}{\scriptsize{} e.g. \citep{Bletzinger01}}\tabularnewline
\hline 
 & {\scriptsize cables} &  & {\scriptsize membrane}\tabularnewline
\hline 
{\scriptsize (a)} & {\scriptsize $\frac{\partial\Pi}{\partial\boldsymbol{x}}=\sum{2wL\nabla L}=\boldsymbol{0}$} & {\scriptsize (b)} & {\scriptsize $\frac{\partial\Pi}{\partial\boldsymbol{x}}=\sum{\nabla S}=\boldsymbol{0}$}\tabularnewline
\hline 
\hline 
 & \multicolumn{3}{c}{\textbf{\scriptsize Surface stress density method}{\scriptsize \citep{Maurin98}}}\tabularnewline
\hline 
 & {\scriptsize membrane} &  & {\scriptsize membrane with cables}\tabularnewline
\hline 
{\scriptsize (c)} & {\scriptsize $\frac{\partial\Pi}{\partial\boldsymbol{x}}=\sum{2wS\nabla S}=\boldsymbol{0}$} & {\scriptsize (d)} & {\scriptsize $\frac{\partial\Pi}{\partial\boldsymbol{x}}=\sum{2wL\nabla L}+\sum{2wS\nabla S}=\boldsymbol{0}$}\tabularnewline
\hline 
\hline 
 & \textbf{\scriptsize Variational Method for Tensegrities}{\scriptsize \citep{kazuma06}} &  & \textbf{\scriptsize Extended Force Density Method}{\scriptsize{} (Proposed)}\tabularnewline
\hline 
 & {\scriptsize cables and struts} &  & {\scriptsize cables and struts}\tabularnewline
\hline 
{\scriptsize (e)} & {\scriptsize $\frac{\partial\Pi}{\partial\boldsymbol{x}}=\sum{kL\nabla L}+\sum{\lambda\nabla L}=\boldsymbol{0}$} & {\scriptsize (f)} & {\scriptsize $\frac{\partial\Pi}{\partial\boldsymbol{x}}=\sum{4wL^{3}\nabla L}+\sum{\lambda\nabla L}=\boldsymbol{0}$}\tabularnewline
\hline 
\end{tabular}
\end{table*}

\section{Conclusions}

In the first part of this work, the extended force density method
was proposed. It enables us to carry out form-finding of prestressed
structures that consist of combinations of both tension and compression
members.

The existence of a variational principle in the FDM was pointed out
and a functional that simply represents the FDM was proposed. Then,
the FDM was extensively redefined by generalizing the formulation
of the functional. Additionally, it was indicated that various functionals
can be selected for form-finding of tension structures. Then, some
form finding analyses of different types of tension structures were
illustrated to show the potential ability of the extended FDM.

In the second part, various form-finding methods were reviewed and
compared in the relation with the extended FDM. By using three types
of expressions such as the principle of virtual work, which can be
commonly found in general problems of statics and are also used in
the description of the extended FDM, the common features and differences
over different form-finding methods were examined.

\section*{Acknowledgments}

This research was partially supported by the Ministry of Education,
Culture, Sports, Science and Technology, Grant-in-Aid for JSPS Fellows,
10J09407, 2010

\appendix

\section{Gradients}

\subsection{Gradient of Linear Element Length}

\label{sub:grad_length}

Let p and q denote two nodes. Let

\noindent 
\begin{equation}
\boldsymbol{p}\equiv\left[\begin{array}{c}
p_{x}\\
p_{y}\\
p_{z}
\end{array}\right]\,\mathrm{,}\,\boldsymbol{q}\equiv\left[\begin{array}{c}
q_{x}\\
q_{y}\\
q_{z}
\end{array}\right]
\end{equation}
represent the Cartesian coordinates of p and q.

The length of the line determined by p and q is given by 
\begin{align}
 & L\left(p_{x},\, p_{y},\, p_{z},\, q_{x},\, q_{y},\, q_{z}\right)\label{eq:length_func}\\
 & \equiv\sqrt{{(p_{x}-q_{x})^{2}+(p_{y}-q_{y})^{2}+(p_{z}-q_{z})^{2}}}.
\end{align}

If the gradient of $L$ is defined by{\scriptsize 
\begin{align}
\nabla L & \equiv\left[\frac{\partial L}{\partial p_{x}},\frac{\partial L}{\partial p_{y}},\frac{\partial L}{\partial p_{z}},\frac{\partial L}{\partial q_{x}},\frac{\partial L}{\partial q_{y}},\frac{\partial L}{\partial q_{z}}\right],
\end{align}
}{\scriptsize \par}

\noindent its components are as follows:{\scriptsize 
\begin{align}
\nabla L & =\left[\frac{p_{x}-q_{x}}{L},\frac{p_{y}-q_{y}}{L},\frac{p_{z}-q_{z}}{L},\frac{q_{x}-p_{x}}{L},\frac{q_{y}-p_{y}}{L},\frac{q_{z}-p_{z}}{L}\right].\label{eq:gradient_L}
\end{align}
}{\scriptsize \par}

\begin{figure}[!tbh]
\centering{}\includegraphics{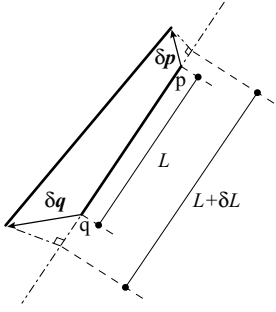}\caption{\label{figLinear element2}Variation of Element Length}
\end{figure}

Let us investigate $\delta L$, i.e.
\begin{equation}
\delta L\equiv\nabla L\cdot\left[\begin{array}{c}
\delta\boldsymbol{p}\\
\delta\boldsymbol{q}
\end{array}\right].
\end{equation}
As shown in Fig. \ref{figLinear element2}, $\delta\boldsymbol{p}$
and $\delta\boldsymbol{q}$ are firstly projected to the line determined
by p and q, then, $\delta L$ is measured on the line. 

\subsection{Gradient of Triangular Element Area}

Let p, q, and r be three vertices. Let 

\textbf{
\begin{equation}
\boldsymbol{p}\equiv\left[\begin{array}{c}
p_{x}\\
p_{y}\\
p_{z}
\end{array}\right],\,\boldsymbol{q}\equiv\left[\begin{array}{c}
q_{x}\\
q_{y}\\
q_{z}
\end{array}\right],\,\boldsymbol{r}\equiv\left[\begin{array}{c}
r_{x}\\
r_{y}\\
r_{z}
\end{array}\right],
\end{equation}
}denote the Cartesian coordinates of p, q, and r.

The area of the triangle determined by p, q, and r is given by 
\begin{align}
S(p_{x},\cdots,r_{z})\equiv & \frac{1}{2}\sqrt{\boldsymbol{N}\cdot\boldsymbol{N}},\\
\left(\boldsymbol{N}\right.\equiv & \left.\left(\boldsymbol{q}-\boldsymbol{p}\right)\times\left(\boldsymbol{r}-\boldsymbol{p}\right)\right).
\end{align}

If the gradient of $S$ is defined by

\begin{equation}
\nabla S\equiv\left[\begin{array}{ccccc}
\frac{\partial S}{\partial p_{x}}, & \frac{\partial S}{\partial p_{y}}, & \frac{\partial S}{\partial p_{z}}, & \cdots & ,\frac{\partial S}{\partial r_{z}}\end{array}\right],
\end{equation}
its components are as follows:{\scriptsize 
\begin{alignat}{1}
\nabla S=\frac{1}{2}\boldsymbol{n}\cdot & \left[(\boldsymbol{r}-\boldsymbol{q})\times\left\{ \left[\begin{array}{c}
1\\
0\\
0
\end{array}\right],\left[\begin{array}{c}
0\\
1\\
0
\end{array}\right],\left[\begin{array}{c}
0\\
0\\
1
\end{array}\right]\right\} \right.\nonumber \\
 & \left.,(\boldsymbol{p}-\boldsymbol{r})\times\left\{ \left[\begin{array}{c}
1\\
0\\
0
\end{array}\right],\left[\begin{array}{c}
0\\
1\\
0
\end{array}\right],\left[\begin{array}{c}
0\\
0\\
1
\end{array}\right]\right\} \right.\label{eq:grad_S}\\
 & \left.,(\boldsymbol{q}-\boldsymbol{p})\times\left\{ \left[\begin{array}{c}
1\\
0\\
0
\end{array}\right],\left[\begin{array}{c}
0\\
1\\
0
\end{array}\right],\left[\begin{array}{c}
0\\
0\\
1
\end{array}\right]\right\} \right],\nonumber 
\end{alignat}
}where $\boldsymbol{n}$ is defined by

\begin{equation}
\boldsymbol{n}\equiv\frac{\boldsymbol{N}}{\left|\boldsymbol{N}\right|}.
\end{equation}

\begin{figure}[!tbh]
\centering{}\includegraphics{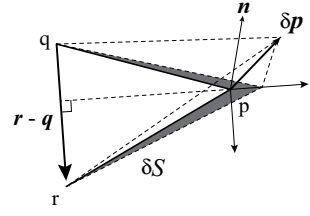}\caption{\label{fig:Triangular-Element2}Variation of Element Area}
\end{figure}

Let us investigate $\delta S$, i.e.{\scriptsize 
\begin{align}
\delta S & =\frac{1}{2}\boldsymbol{n}\cdot\left((\boldsymbol{r-q})\times\delta\boldsymbol{p}+(\boldsymbol{p}-\boldsymbol{r})\times\delta\boldsymbol{q}+(\boldsymbol{q}-\boldsymbol{p})\times\delta\boldsymbol{r}\right).\label{eq:deltaS}
\end{align}
}With respect to $\delta\boldsymbol{p}$, for example, when $\delta\boldsymbol{p}$
is orthogonal to the element, $\left(\boldsymbol{r}-\boldsymbol{q}\right)\times\delta\boldsymbol{p}$
becomes orthogonal to $\boldsymbol{n}$, then $\delta S$ vanishes
(see Fig. \ref{fig:Triangular-Element2}). On the other hand, when
$\delta\boldsymbol{p}$ is parallel to the opposite side, $\left(\boldsymbol{r}-\boldsymbol{q}\right)\times\delta\boldsymbol{p}$
vanishes, then $\delta S$ vanishes. Therefore, only the component
of $\delta\boldsymbol{p}$ which is parallel to the perpendicular
line from p to the opposite side can produce $\delta S$. In other
words, $\delta S$ is measured on the plane determined by p, q, and
r.

\section{Some Remarks of Surface Area}

\subsection{Minimal Surfaces and Uniform Stress Surfaces}

The surface area of a surface is given by

\begin{equation}
a=\int_{a}\mathrm{da}.\label{eq:a}
\end{equation}
Here, $\mathrm{da}$ is called area element and defined by
\begin{equation}
\mathrm{da}\equiv\sqrt{\mathrm{det}g_{ij}}d\theta^{1}d\theta^{2},\label{eq:da}
\end{equation}
where $g_{ij}$ and $\left\{ \theta^{i},\theta^{j}\right\} $ are
the Riemannian Metric and the local coordinate on the surface respectively.

Using Eq. \eqref{eq:da}, the variation of the surface area, $\delta a$,
can be calculated and the result is as follows: 
\begin{equation}
\delta a=\frac{1}{2}\int_{a}g^{ij}\delta g_{ij}\sqrt{\mathrm{det}g_{ij}}d\theta^{1}d\theta^{2},
\end{equation}
\begin{equation}
\therefore\delta a=\frac{1}{2}\int_{a}g^{ij}\delta g_{ij}\mathrm{da},\label{eq:var_area-2}
\end{equation}
where $g^{ij}$ is the inverse matrix of $g_{ij}$.

By the way, on a membrane, the 2nd \textit{Piola-Kirchhoff} stress
tensor and the \textit{Green-Lagrange} strain tensor are defined by
\begin{equation}
\boldsymbol{S}\equiv\frac{\sqrt{\mathrm{det}g_{ij}}}{\sqrt{\mathrm{det}\bar{g}_{ij}}}T_{\cdot k}^{i}g^{kj}\bar{\boldsymbol{g}}_{i}\otimes\bar{\boldsymbol{g}}_{j}\,,\,\boldsymbol{E}\equiv\frac{1}{2}(g_{ij}-\bar{g}_{ij})\bar{\boldsymbol{g}}^{i}\otimes\bar{\boldsymbol{g}}^{j},
\end{equation}
where $T_{\cdot k}^{i}$ are the components of the Cauchy stress tensor.
In addition, $\bar{\boldsymbol{g}}_{i},\,\bar{\boldsymbol{g}}^{i},\,\bar{g}_{ij}$
are the dual bases and the Riemannian metric defined on a reference
configuration.

Then, the \textbf{Principle of Virtual Work} for membranes is expressed
as: 
\begin{equation}
\delta w=\int_{\bar{a}}t\boldsymbol{S}:\delta\boldsymbol{E}\mathrm{d\bar{a}}\label{eq:pvw_membrane-3}
\end{equation}
where $\mathrm{d\bar{a}},\,\bar{a}$ are related to the reference
configuration, and $t$ denotes the thickness. Eq. \eqref{eq:pvw_membrane-3}
reduces to the following form: 
\begin{equation}
\delta w=\int_{a}tT_{\cdot k}^{i}g^{kj}\delta g_{ij}\mathrm{da},
\end{equation}
which does not depend on the reference configuration.

Because Eq. \eqref{eq:var_area-2} can be transformed into the following
form: 
\begin{equation}
\delta a=\int_{a}\delta_{\cdot k}^{i}g^{kj}\delta g_{ij}\mathrm{da},
\end{equation}
when $t$ and $T_{\cdot k}^{i}$ are uniform on the surface and when
$T_{\cdot k}^{i}=\hat{\sigma}\delta_{\cdot k}^{i}$, where $\hat{\sigma}$
is also uniform, then 
\begin{equation}
\delta w=t\hat{\sigma}\delta a\,\,\,\,\therefore\delta w=0\Leftrightarrow\delta a=0,
\end{equation}
which is a simple demonstration of the essential identity of uniform
stress surfaces and minimal surfaces.

\subsection{Galerkin Method for Minimal Surface}

When the form of a surface is represented by $n$-independent parameters
such as $\boldsymbol{x}=\left[\begin{array}{ccc}
x_{1} & \cdots & x_{n}\end{array}\right]^{T}$, an approximation of 
\begin{equation}
\delta a=\int_{a}g^{ij}\delta g_{ij}\mathrm{da}=0
\end{equation}
can be obtained by the \textit{Galerkin method} and it is as follows:
\begin{equation}
\delta\tilde{a}=\left(\int_{a}g^{ij}\nabla g_{ij}\mathrm{da}\right)\cdot\delta\boldsymbol{x}=0,
\end{equation}
where $\nabla$ is the gradient operator defined by
\begin{equation}
\nabla f\equiv\left[\begin{array}{ccc}
\frac{\partial f}{\partial x_{1}} & \cdots & \frac{\partial f}{\partial x_{n}}\end{array}\right]
\end{equation}
and $\delta\boldsymbol{x}$ is the variation of $\boldsymbol{x}$,
or, just an arbitrary column vector.

When the form is discretized by $m$ elements, the integral can be
divided into $m$ independent integrals. Hence 
\begin{equation}
\delta\tilde{a}=\left(\sum_{j}\left.\int_{a}g^{\alpha\beta}\nabla g_{\alpha\beta}\mathrm{da}\right|_{j}\right)\cdot\delta\boldsymbol{x},
\end{equation}
where $j$ is the index of each element.

In each element, remembering the relation of 
\begin{equation}
\left.\int_{a}g^{\alpha\beta}\delta g_{\alpha\beta}\mathrm{da}\right|_{j}=\left.\delta\int_{a}\mathrm{da}\right|_{j},
\end{equation}
the following transformation is also correct:
\begin{equation}
\left.\int_{a}g^{\alpha\beta}\nabla g_{\alpha\beta}\mathrm{da}\right|_{j}=\left.\nabla\int_{a}\mathrm{da}\right|_{j},
\end{equation}
due to the fact that $\delta$ symbol is originally defined by $\frac{\partial}{\partial\epsilon}$
when $\epsilon$ is the assigned one-parameter to represent the change
of the form.

Therefore, when $S_{j}$ is defined as a function to give $j$-th
element area, i.e. 
\begin{equation}
S_{j}\equiv\left.\int_{a}\mathrm{da}\right|_{j},
\end{equation}
then
\begin{equation}
\delta\tilde{a}=0\Leftrightarrow\left(\sum_{j}\nabla S_{j}\right)\cdot\delta\boldsymbol{x}=0\Leftrightarrow\sum_{j}\nabla S_{j}=\boldsymbol{0},
\end{equation}
which is the stationary condition of 
\begin{equation}
\Pi\left(\boldsymbol{x}\right)=\sum_{j}S_{j}(\boldsymbol{x}).
\end{equation}

\end{document}